\documentclass[10pt]{article}

\usepackage[a4paper,margin=1in]{geometry}
\usepackage[T1]{fontenc}
\usepackage[utf8]{inputenc}
\usepackage{lmodern}
\usepackage{microtype}
\usepackage{amsmath,amssymb,mathtools,bm}
\usepackage{graphicx}
\usepackage{xcolor}
\usepackage[numbers,sort&compress]{natbib}

\definecolor{darkblue}{rgb}{0,0,0.6}
\definecolor{darkred}{rgb}{0.6,0,0}
\usepackage[colorlinks=true,urlcolor=darkblue, citecolor=darkblue, linkcolor=darkred, hyperfootnotes=false]{hyperref}
\usepackage[nameinlink,noabbrev]{cleveref}

\def\dd{\mathrm{d}}
\def\rb{\bar\rho}

\def\rhom{\rho_0}

\title{\bfseries Transverse geometry reshapes current and tracer fluctuation amplitudes in quasi-one-dimensional single files}
\author{}
\date{}

\begin{document}

\begin{center}
    {\large \textbf{Transverse geometry reshapes current and tracer fluctuation amplitudes in quasi-one-dimensional single files} }
    
    \medskip

    {Olivier Bénichou${}^\star$ and Aurélien Grabsch}\\
    \textit{Sorbonne Université, CNRS, Laboratoire de Physique Théorique de la\\
Matière Condensée (LPTMC), 4 Place Jussieu, 75005 Paris, France}\\[0.1cm]
    $\star$ \texttt{benichou@lptmc.jussieu.fr}
\end{center}

\begin{abstract}
Single-file transport means no overtaking: particles move in a narrow channel
while preserving their longitudinal order. This simple constraint has
profound dynamical consequences, most notably tracer subdiffusion, and has
made single-file transport a paradigmatic form of confined many-body motion,
observed from molecular transport in zeolites to single-file diffusion of
colloids in narrow channels. A common modelling approach assumes that,
once overtaking is suppressed, finite-width effects can be discarded or
represented by an effective particle size in a strictly one-dimensional
model. Starting from the Brownian dynamics in the full confined geometry, we
instead derive the exact large-scale one-dimensional
fluctuating-hydrodynamic equation governing the coarse-grained line density.
Its transport coefficients are fixed by the confined equilibrium equation of
state, through which the transverse geometry remains encoded in the
one-dimensional description. Consequently, transverse geometry leaves the
$t^{1/2}$ single-file scaling unchanged but can qualitatively reshape the
density dependence of the current and tracer fluctuation amplitudes. In the
minimal hard-core setting, this yields a collective diffusivity that
can become non-monotonic in density for sufficiently wide no-passing
channels. This geometric non-monotonicity propagates to exact large-scale
predictions for the amplitudes of integrated-current fluctuations and
tracer-displacement fluctuations. The effect is robust to the interaction
potential, channel geometry, initial preparation and microscopic dynamics.
Quasi-one-dimensional single-file transport therefore defines a distinct
regime in which forbidding overtaking does not erase geometry from collective
transport.
\end{abstract}

When particles are confined so tightly that they cannot pass each other, their dynamics enters the single-file regime. The preservation of order then has direct dynamical consequences, most notably the subdiffusive spreading of a tagged particle, and has made single-file transport a paradigmatic form of confined many-body motion \cite{Harris:1965,Levitt:1973,Kollmann:2003,Barkai:2009,Hegde:2014,Metzler:2014,Krapivsky:2014,Krapivsky:2015a,Sadhu:2015,Poncet:2021,Grabsch:2024a,Taloni:2017,Solano:2025}. This regime has been observed from molecular transport in zeolites to colloidal particles in quasi-one-dimensional channels \cite{Hahn:1996,Kukla:1996,Wei:2000,Lutz:2004,Lin:2005}.

In experiments, however, single-file transport is never strictly one-dimensional. Particles are confined in pores, channels or trenches with a finite transverse section \cite{Karger:2012,Zwanzig:1992,Burada:2009,Franosch:2012,Benichou:2018,Kavokine:2021}. The width may be small enough to forbid overtaking, but not small enough to freeze the transverse degrees of freedom. This is the quasi-one-dimensional single-file regime considered here. The central question is then whether this residual geometry only produces microscopic corrections or alters the large-scale transport laws themselves.

This question has not, to our knowledge, been addressed directly. In strictly one-dimensional systems, both the $t^{1/2}$ scaling of the tracer mean-squared displacement and the dependence of its amplitude on density and equilibrium compressibility are well established~\cite{Kollmann:2003,Hegde:2014}. The specific role of finite width while overtaking remains forbidden, however, appears in the literature only indirectly, through three distinct lines of work.

First, many studies motivated by transport in narrow or quasi-one-dimensional confinements nevertheless formulate the problem directly in strictly one-dimensional terms~\cite{Kollmann:2003,Lin:2005,Hegde:2014,Lizana:2010,Krapivsky:2014,Taloni:2017,Ryabov:2019,Poncet:2021,Grabsch:2022,Vorac:2023,Schweers:2023,Saha:2026b}. These works do not explicitly claim an exact equivalence between finite-width and strictly one-dimensional systems. By eliminating the transverse degrees of freedom from the outset, however, these descriptions do not allow finite channel width to enter the constitutive transport coefficients within the no-passing regime.

Second, studies that explicitly examine the role of channel width have mainly focused on the breakdown of the single-file regime. Once the confinement becomes wide enough to allow particles to bypass one another, tracer motion crosses over at long times from single-file subdiffusion to normal diffusion. This crossover, and its dependence on the bypassing rate, have been studied extensively~\cite{Mon:2002,Lucena:2012,Richardson:2025}. These works therefore address how finite width destroys single-file behaviour, rather than how it affects large-scale transport while particle order remains strictly preserved.

Third, a few studies have explicitly considered finite-width systems within the no-passing regime by mapping the confined system onto a strictly one-dimensional hard-rod model with an effective particle diameter or occupied volume fraction~\cite{Lin:2005,Mon:2005,Zarif:2020}. These approaches allow the channel width to affect transport, but represent
its effect through an effective one-dimensional particle size or occupied
volume fraction. As we show below, these reductions fail even qualitatively to capture the density dependence of the current and tracer fluctuation amplitudes in finite-width channels.

\begin{figure}[t]
\centering
\includegraphics[width=\linewidth]{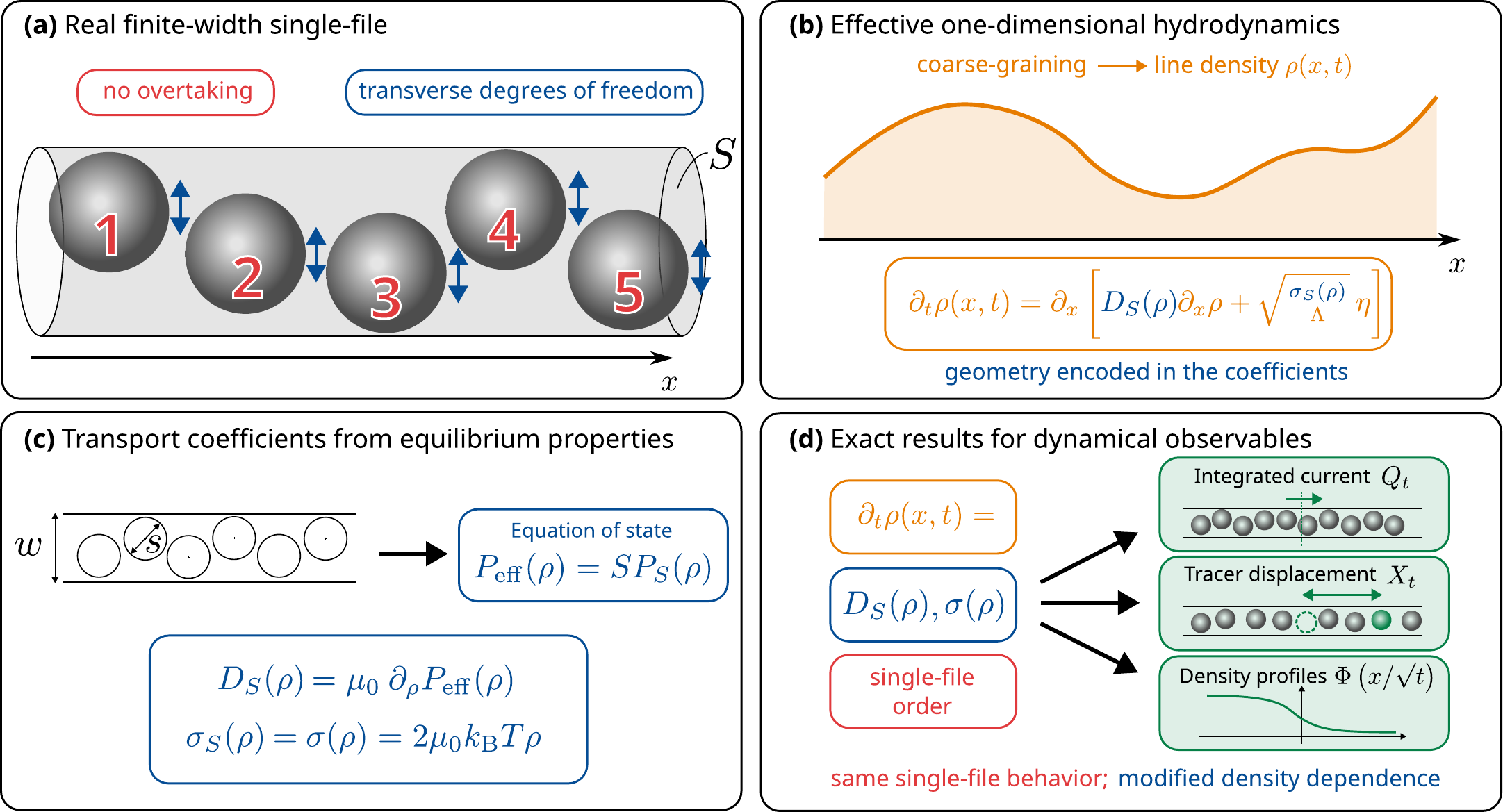}
\caption{\textbf{Geometry enters single-file dynamics through the transport coefficients.}
\textbf{(a)} Brownian particles diffuse in a channel of transverse section $S$. Their longitudinal order is conserved, which forbids overtaking, while transverse exploration remains allowed. This allows one to define a numbering of the particles which remains the same at all times.
\textbf{(b)} Coarse-graining over the transverse coordinates defines the line density $\rho(x,t)$. On large scales, $\rho(x,t)$ obeys an exact one-dimensional fluctuating hydrodynamics. The field is one-dimensional, and all microscopic information enters through the transport coefficients $D_S(\rho)$ and $\sigma_S(\rho)$.
\textbf{(c)} The confined equilibrium problem determines the effective pressure $P_{\rm eff}(\rho)=S P_S(\rho)$. This pressure fixes the geometry-dependent collective diffusivity $D_S(\rho)=\mu_0\partial_\rho P_{\rm eff}(\rho)$, whereas the line mobility is $\sigma(\rho)=2\mu_0 k_{\rm B}T\rho$.
\textbf{(d)} The transport coefficients, together with the preserved single-file order, determine the integrated current $Q_t$, the tracer displacement $X_t$, and the bath--tracer density profile $\Phi(x/\sqrt t)$. Compared to the purely one-dimensional case, the observables retain their single-file scaling, but acquire a geometry-dependent density dependence.}
\label{fig:system}
\end{figure}

Here we show that the transverse structure remains encoded in the large-scale transport observables considered here. Starting from the Brownian dynamics in the full confined geometry, we derive an exact large-scale one-dimensional fluctuating hydrodynamics for the longitudinal density\footnote{Effective one-dimensional reductions of confined diffusion have a long history, notably through Fick--Jacobs and related projection methods~\cite{Kalinay:2007,Kalinay:2008}.}. Its transport coefficients are determined by the equation of state of the confined ordered fluid, which we compute using a transfer-matrix method, while the preservation of particle order gives tracer motion and current fluctuations their single-file character. This yields exact large-scale predictions for the collective diffusivity, current fluctuations, tracer fluctuations and the associated bath--tracer density profile, without any phenomenological input. For hard-core particles, the collective diffusivity can become non-monotonic in density in sufficiently wide no-passing channels, producing a second maximum in the amplitude of the current fluctuations and a plateau in that of the tracer fluctuations. This revises the conventional picture of single-file transport: the single-file constraint fixes the universal $t^{1/2}$ scaling, whereas the equation of state of the confined fluid fixes the density and width dependence of the fluctuation amplitudes and can reshape it qualitatively. The resulting finite-width effects are robust to the interaction potential, channel geometry, initial preparation and microscopic particle dynamics, whether overdamped or underdamped. The logic of the construction is summarized in Fig.~\ref{fig:system}.

\textit{Fluctuating hydrodynamic description.---}
We consider \(N\) Brownian particles of diameter \(s\) in a $d$-dimensional channel of length \(L\) and finite transverse section \(S\). The longitudinal coordinate is denoted by \(x\), and the $(d-1)$-dimensional transverse coordinate by \({\bf y}\).
The microscopic positions \({\bf r}_n=(x_n,{\bf y}_n)\) evolve according to
\begin{equation}
\frac{{\rm d}{\bf r}_n}{{\rm d}t}
=-\mu_0 \sum_m \nabla V({\bf r}_n-{\bf r}_{m})
+\sqrt{2\mu_0 k_{\rm B}T}\,{\boldsymbol\eta}_n(t)
\:,
\label{eq:langevin_channel}
\end{equation}
with reflecting boundary conditions at the channel walls. Here ${\boldsymbol\eta}_n$ is a Gaussian white noise, $\mu_0$ is the microscopic mobility and $V$ includes the pair interaction and the hard-core constraint. We assume that the pair interaction decays faster than $|x|^{-1}$ so that the total energy is extensive and standard thermodynamics can be defined. This ensures that the dynamics remains diffusive at large scales\footnote{See for instance Ref.~\cite{Dandekar:2023} for longer range pair potentials which yield a non-diffusive behaviour}.

The relevant density is the line density \(\bar\rho=N/L\), or, locally, the microscopic density
\begin{equation}
    \label{eq:defMicroDens}
    \rhom(\mathbf{r},t) =
    \sum_n \delta^{(d)} \left( \mathbf{r} - \mathbf{r}_n(t) \right)
    \:,
\end{equation}
where $\delta^{(d)}$ is the $d$-dimensional delta function. The time evolution of this density is given by the Dean-Kawasaki equation~\cite{Kawasaki:1994,Dean:1996,Illien:2024}, which is equivalent to the set of Langevin equations~\eqref{eq:langevin_channel}. However, this equation being non-local and stochastic, it remains mostly formal and cannot be analyzed for the hard-core interaction considered here~\cite{Demery:2014,Illien:2024}. We thus instead consider the coarse-grained longitudinal density \(\rho(x,t)\), defined as
\begin{equation}
    \label{eq:defLongDens}
    \rho(x,t) =
    \int_{-\infty}^\infty \dd x' \int_S \dd^{d-1} {\mathrm{y'}} \:
    \phi(\Lambda x-x') 
    \rhom \big( (x',y'), \Lambda^2 t \big)
    \:,
\end{equation}
where $\Lambda$ is a rescaling factor from the microscopic to the macroscopic description, and $\phi$ is a window function centered around the origin, with $\int_{-\infty}^\infty \phi(x) \dd x= 1$. 
The diffusive rescaling \(x\to \Lambda x\), \(t\to \Lambda^2 t\) is the one appropriate for the large-scale density dynamics.

At large scales $\Lambda \gg 1$, we show in the Supplementary Information (SI) by coarse-graining the Dean-Kawasaki equation, adapting the methods used for unconfined geometries~\cite{Das:2013,Saha:2026} to the channel case, that the longitudinal density obeys an exact fluctuating hydrodynamic equation of the form~\cite{Spohn:1991,Bertini:2015,Derrida:2007}
\begin{equation}
\partial_t\rho=\partial_x\!\left[D_S(\rho)\partial_x\rho+
\sqrt{\frac{\sigma_S(\rho)}{\Lambda}}\,\eta\right],
\label{eq:fh}
\end{equation}
where \(D_S(\rho)\) is the collective diffusivity, \(\sigma_S(\rho)\) the mobility of the line density, and \(\eta(x,t)\) a Gaussian white noise in space and time, 
\begin{equation}
    \langle\eta(x,t)\eta(x',t')\rangle=\delta(x-x')\delta(t-t') 
\:.
\end{equation}
Compared to the Dean-Kawasaki equation, the main advantage of the macroscopic description is that the evolution equation~\eqref{eq:fh} is local, with a weak noise, due to the rescaling factor $\Lambda \gg 1$. This equation is exact for the interaction potentials considered here and under the assumption that, at large scale, the system can be considered to be locally at equilibrium. This is a standard and rather expected assumption, since the line density $\rho$ is the only conserved field it is expected to relax much slower than all other degrees of freedom\footnote{In the diffusive systems considered here, the relaxation to equilibrium of the transverse degrees of freedom is efficient, and does not involve anomalously long relaxation times as reported in ballistic systems~\cite{Mayo:2022}}. The local equilibrium property however remains difficult to prove rigorously beyond schematic models like the exclusion process~\cite{Kipnis:1989}.
The dependence on the transverse section \(S\) is entirely contained in the transport coefficients.
At large scales, the $d$-dimensional confined system is therefore reduced to
a one-dimensional fluctuating hydrodynamics for the conserved line density.
Crucially, this reduction does not replace the transverse geometry by a single
effective particle size: the full density dependence of the transport
coefficients retains the  structure of the confined fluid.

The transport coefficients can be shown to be related to the equilibrium thermodynamics of the confined fluid by following the framework developed for interacting Brownian
particles with general interactions~\cite{Grabsch:2025b,Grabsch:2026}. Their expressions follow directly 
from the microscopic coarse-graining given in the SI. Alternatively, they can also be
understood from linear response and fluctuation--dissipation, as detailed now. We first
determine
the mobility $\sigma_S(\rho)$ as the linear response of the line current $j$ to the application of a small uniform external force $F_0$, \(\langle j\rangle=\sigma_S(\rho)F_0/(2k_{\rm B}T)\). The application of this force along the channel amounts to a change of frame with velocity \(\mu_0F_0\), hence
\begin{equation}
  \sigma(\rho)=2\mu_0 k_{\rm B}T\rho
  \:,
\label{eq:mobility}
\end{equation}
The index $S$ has been dropped because the line mobility is independent of the transverse section.
The diffusivity then follows from the fluctuation--dissipation relation which relates $D_S$ and $\sigma$ \cite{Bertini:2015,
Derrida:2025a},
\begin{equation}
D_S(\rho)=\mu_0\partial_\rho P_{\rm eff}(\rho)
\:,
\label{eq:diffusivity_pressure}
\end{equation}
where \(P_{\rm eff}=S P_S\) is the effective one-dimensional pressure, and \(P_S\) is the longitudinal pressure of the fluid confined in the section \(S\). The dynamical problem is thus controlled by a single  equilibrium quantity: the equation of state of the ordered fluid in the channel.
Note that the fluctuating hydrodynamic equation~\eqref{eq:fh} holds for any interaction and channel width, regardless of the single-file constraint.

\textit{Confined equation of state.---}
We now consider the single-file regime in which the width $w$ of the channel is smaller than $2s$. To determine the effective pressure $P_{\rm eff}$, we further assume that the interactions occur only between nearest neighbours. For purely hard-core interactions, this is the case for width $w < (1+\sqrt{3}/2)s\simeq 1.866\,s$.
In this regime, the non-overtaking constraint together with the nearest-neighbour interaction makes the partition function computable by a transfer matrix method \cite{Kofke:1993,Montero:2023,Franosch:2024,Montero:2025}.
Introducing the grand-canonical partition function \(\Xi(\beta,\mu;x,{\bf y})\) for a file on the positive axis ending with a last particle at longitudinal position \(x\) and transverse coordinate \({\bf y}\), one obtains the recursion relation
\begin{equation}
\Xi(\beta,\mu; x,{\bf y})-\frac{e^{\beta\mu}}{\ell_0^d}
\int_0^x{\rm d}x'\int_S{\rm d}^{d-1}{\bf y}'\,
e^{-\beta V({\bf r}-{\bf r}')}\Xi(\beta,\mu;x',{\bf y}')=1
\:.
\label{eq:Xi_equation}
\end{equation}
At large \(x\), the growth of \(\Xi\) defines the grand potential, and thus the pressure $P_{\mathrm{eff}}$,
\(\Xi(\beta,\mu;x,{\bf y})\sim e^{\beta P_{\rm eff}x}\phi({\bf y})\), where \(\phi\) is the dominant transverse eigenmode. Taking the Laplace transform in \(x\) then gives an exact transverse eigenvalue problem
\begin{equation}
\ell_0^d e^{-\beta\mu}\phi({\bf y})
=\int_S{\rm d}^{d-1}{\bf y}'\,
\widehat V(\beta;\beta P_{\rm eff},{\bf y}-{\bf y}')\phi({\bf y}')
\:,
\qquad
\partial_\mu P_{\rm eff}=\rho
\:,
\label{eq:eos_eigenproblem}
\end{equation}
where
\begin{equation}
\widehat V(\beta;u,{\bf y})=\int_0^\infty{\rm d}x\,e^{-ux}
\exp[-\beta V(\sqrt{x^2+|{\bf y}|^2})]
\:.
\label{eq:Vhat}
\end{equation}
Thus the pressure of the confined file is selected by the dominant transverse eigenvalue, and the longitudinal equation of state follows from the Legendre transformation to the canonical ensemble \(\partial_\mu P_{\rm eff}=\rho\).
The eigenvalue problem can be solved numerically, as shown in the SI along details on the Legendre transform and derivation of the eigenvalue equation~\eqref{eq:eos_eigenproblem}.

For purely hard particles, this transverse eigenvalue problem takes a particularly simple form. The Boltzmann weight is
\begin{equation}
   e^{-\beta V({\bf r})}=\Theta(|{\bf r}|^2-s^2)
   \:, 
\end{equation}
so that, 
\begin{equation}
\widehat V(u,y)=\frac{1}{u}\exp[-u\sqrt{s^2-y^2}]
\:.
\label{eq:hard_kernel}
\end{equation}
The transverse offset therefore changes the longitudinal excluded distance. Solving the exact eigenvalue problem~(\ref{eq:eos_eigenproblem}) gives \(P_{\rm eff}(\rho)\), and Eq.~(\ref{eq:diffusivity_pressure}) gives the collective diffusivity. Although particle order is conserved exactly as in a strictly
one-dimensional hard-rod system, the pressure and the collective diffusivity
\(D_S(\rho)\) differ from their strictly one-dimensional counterparts. In particular, the collective diffusivity can become non-monotonic in density for sufficiently wide channels (while still ensuring the single-file constraint), as shown in the SI.

\begin{figure}[t]
\centering
\includegraphics[width=0.9\textwidth]{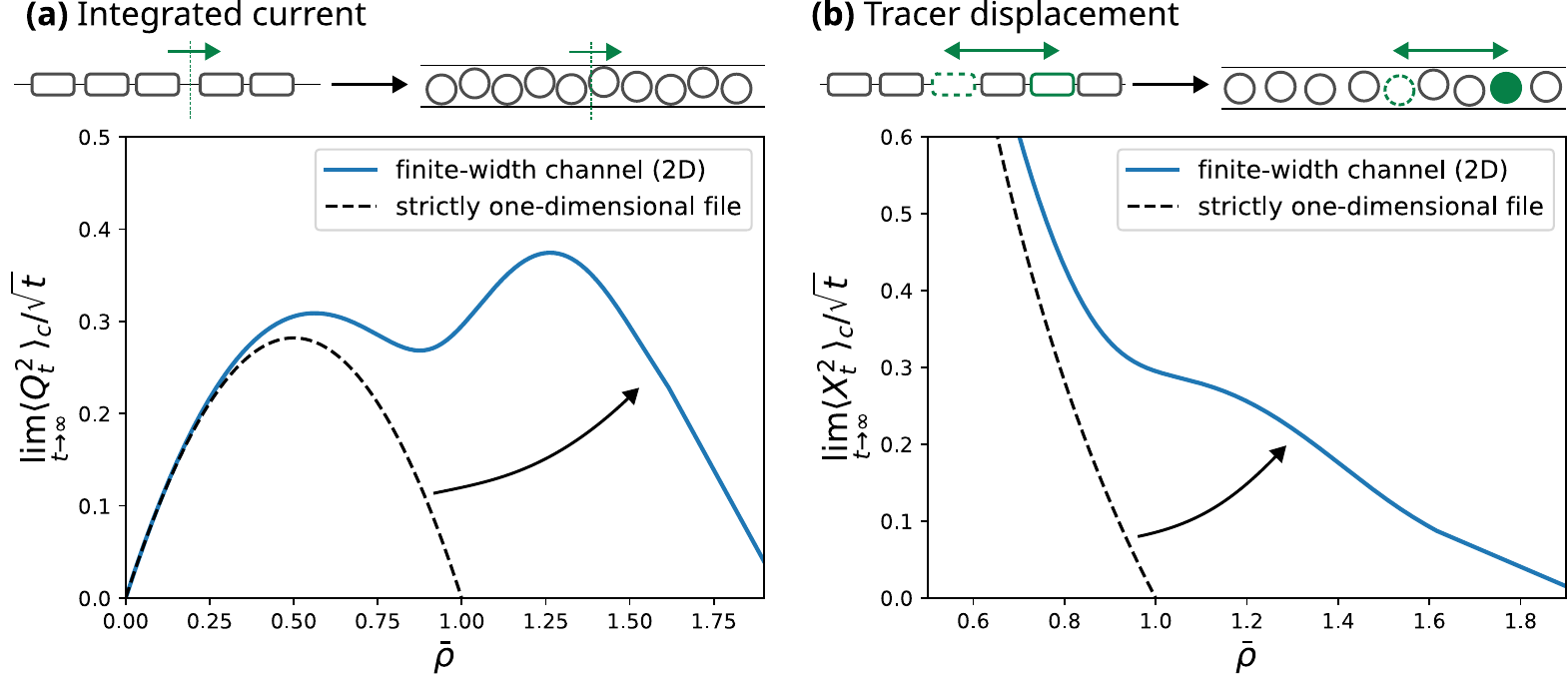}
\caption{\textbf{Finite width reshapes the density dependence of dynamical observables.} 
\textbf{(a)} Fluctuations of the integrated current. 
\textbf{(b)} Fluctuations of the displacement of a tracer. The strictly one-dimensional hard-rod description does not reproduce the qualitatively different density dependence found in a finite-width two-dimensional channel of hard disks. The plots are for particles of size $s=1$ in a channel of width $w=1.86$.}
\label{fig:effect0}
\end{figure}

\textit{Application to current and tracer fluctuations.---}
Having determined \(D_S(\rho)\) and \(\sigma(\rho)\), Eq.~\eqref{eq:fh}, together with the initial condition, gives the exact
leading large-scale behaviour of the single-file observables. Single-file systems are known to retain a lasting memory of their initial preparation in dynamical observables~\cite{Leibovich:2013}.
The variance of the integrated current $Q_t$ through a section of the channel
is obtained by expanding Eq.~\eqref{eq:fh} around the uniform density
$\bar\rho$~\cite{Krapivsky:2012}, 
\begin{equation}
\langle Q_t^2\rangle_{\rm c}
\simeq
\frac{\sigma(\rb)}{\sqrt{\pi D_S(\rb)}}\,\sqrt{t}
\label{eq:current_variance_annealed}
\end{equation}
for an annealed preparation, in which the initial density is sampled from
equilibrium. For a quenched preparation, in which the initial density is
fixed, this variance is smaller by a factor $\sqrt{2}$.
Similarly, it can be shown that the tracer displacement remains subdiffusive, with~\cite{Kollmann:2003,Krapivsky:2014,Krapivsky:2015a}
\begin{equation}
\langle X_t^2\rangle_{\rm c}
\simeq
\frac{\sigma(\rb)}{\rb^2\sqrt{\pi D_S(\rb)}}\,\sqrt{t}
\label{eq:tracer_variance}
\end{equation}
in the annealed case. For a quenched preparation, the tracer variance is again smaller by a factor
$\sqrt{2}$. These formulae are the usual single-file results \cite{Krapivsky:2014,Krapivsky:2015a,Sadhu:2015,Grabsch:2024b}, which we rederive in the SI from the evolution equation~\eqref{eq:fh}. Importantly, here the transverse structure of the system is encoded in  the geometry-dependent collective diffusivity~\eqref{eq:diffusivity_pressure} obtained from the equation of state~\eqref{eq:eos_eigenproblem}. Figure~\ref{fig:effect0} shows that this geometry dependence is not a small
correction: the density dependence of the current and tracer fluctuations is reshaped by the channel geometry.

The origin of this finite-width dependence is the competition between
longitudinal crowding and transverse organization. In a strictly
one-dimensional hard-rod file, increasing the density only reduces the
available longitudinal free length. In a finite-width file, compression
can also induce transverse rearrangements: larger transverse offsets reduce
the longitudinal distance at contact between neighbouring particles. While
this reorganization develops, part of the density increase is accommodated
by this decrease of the contact distance; once the transverse packing is
nearly saturated, further compression again acts mainly on the longitudinal
free gaps. This crossover produces the non-monotonic collective diffusivity.
Through Eq.~\eqref{eq:diffusivity_pressure}, this change in equilibrium
packing is converted directly into a qualitative change of the transport
laws.

\begin{figure}[t]
\centering
\includegraphics[width=0.9\textwidth]{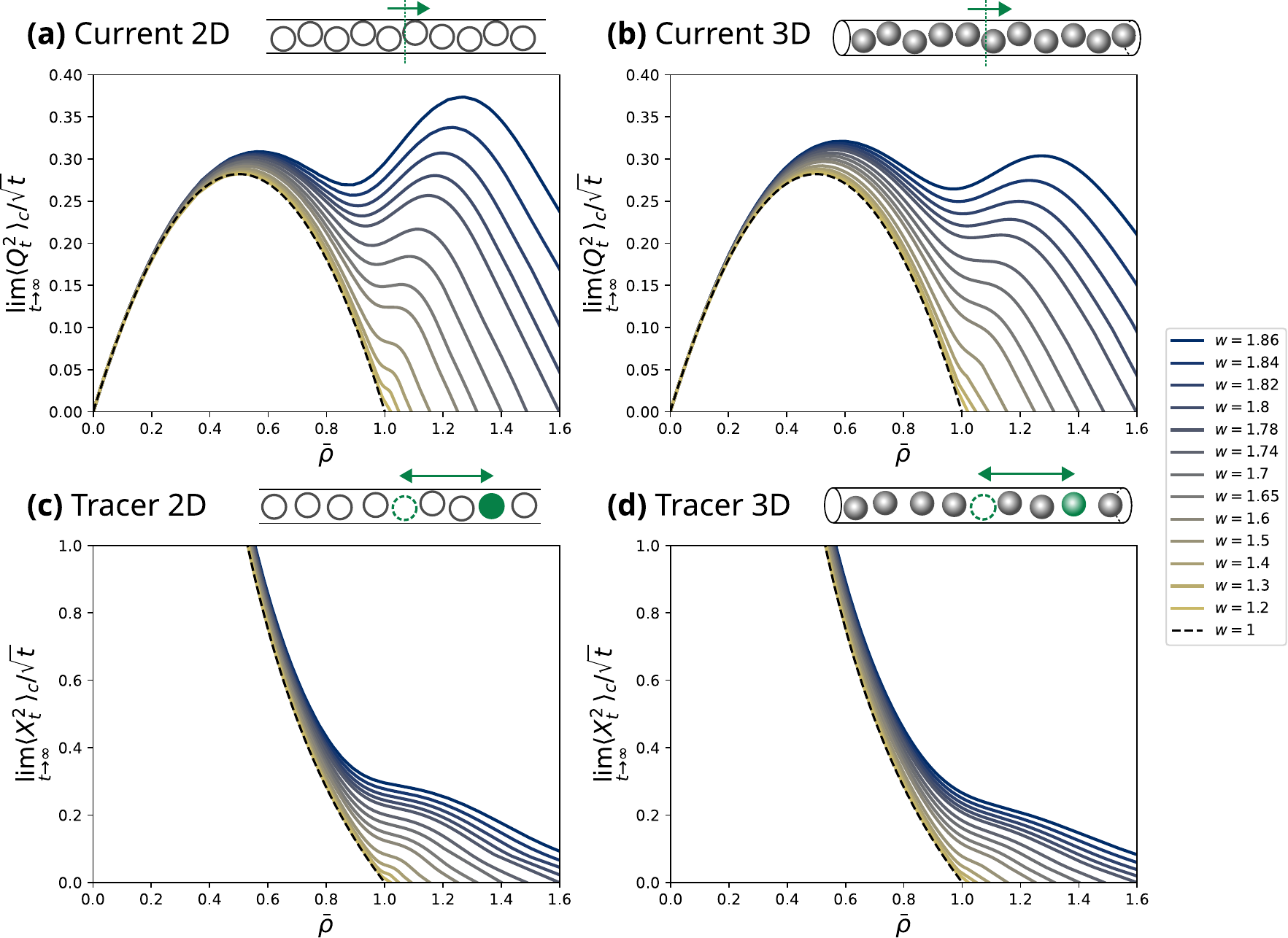}
\caption{\textbf{Robustness with respect to channel geometry.} The effect is robust to the dimension and transverse size $w$ of the channels. The plots are for hard-core particles of size $s=1$ with an annealed initial condition. The strictly one-dimensional case corresponds to $w=s=1$.}
\label{fig:robustness}
\end{figure}

The current fluctuations make the consequence of this transverse reorganization especially transparent. In a strictly one-dimensional hard-rod file, the density dependence of \(\langle Q_t^2\rangle_{\rm c}/\sqrt{t}\) is fixed by a simple free-volume law, and the density of maximal fluctuations follows from that one-dimensional packing constraint. In a finite-width file, this optimum is no longer a one-dimensional property: a second local maximum of fluctuations appears at a higher density as the width of the channel is increased, as shown in Fig.~\ref{fig:robustness}(a) and~(b). For sufficiently wide channels, this second maximum becomes the true maximum of the current fluctuations. This unexpected effect cannot be accounted for by a simple rescaling of the density or of the size of the particles, as discussed below.  Geometry therefore gives rise to a behaviour that cannot be reproduced by a
strictly one-dimensional hard-rod description. For the tracer-displacement fluctuations, shown in Fig.~\ref{fig:robustness}(c) and~(d), a plateau emerges at densities $\bar\rho s >1$, which does not exist in a one-dimensional system.
This plateau provides another signature of the finite-width effect.

Beyond the amplitudes of current and tracer fluctuations, the theory also gives access to their spatial organization. We determine, in the tracer frame, the density response associated with a tracer displacement. This response does not resolve the microscopic transverse configurations themselves; rather, it gives their large-scale longitudinal signature. The relevant object is the connected bath--tracer correlation profile~\cite{Poncet:2021}, which has the scaling form
\begin{equation}
\langle \rho(X_t+x,t)X_t\rangle_{\rm c}
\simeq
\Phi\!\left(\frac{x}{\sqrt{t}}\right).
\label{eq:density_profile_scaling}
\end{equation}
This profile is the large-scale density response associated with a tracer displacement: it correlates a spontaneous tracer displacement with the excess density around it. Equivalently, it quantifies the accumulation of particles in front of a tracer that has moved in one direction and the depletion behind it.
For an equilibrium annealed preparation,
\begin{equation}
\Phi(z)=
\frac{\sigma(\rb)}{4\rb D_S(\rb)}
\,{\rm erfc}\!\left(\frac{z}{\sqrt{4D_S(\rb)}}\right),
\qquad z>0,
\label{eq:density_profile}
\end{equation}
with \(\Phi(-z)=-\Phi(z)\). For a quenched initial condition, the same formula holds with $z$ replaced by $z/\sqrt{2}$. The transverse geometry therefore does not only change the amplitude of tracer fluctuations; it also fixes the spatial structure of the bath rearrangement that produces them. In this sense, the profile provides a direct large-scale signature of the finite-width organization of the file.
Thus the exactness is not limited to the transport coefficients: once the confined equation of state is known, the large-scale current fluctuations, tracer fluctuations and bath-density response follow without further microscopic input.

\begin{figure}[t]
\centering
\includegraphics[width=0.8\textwidth]{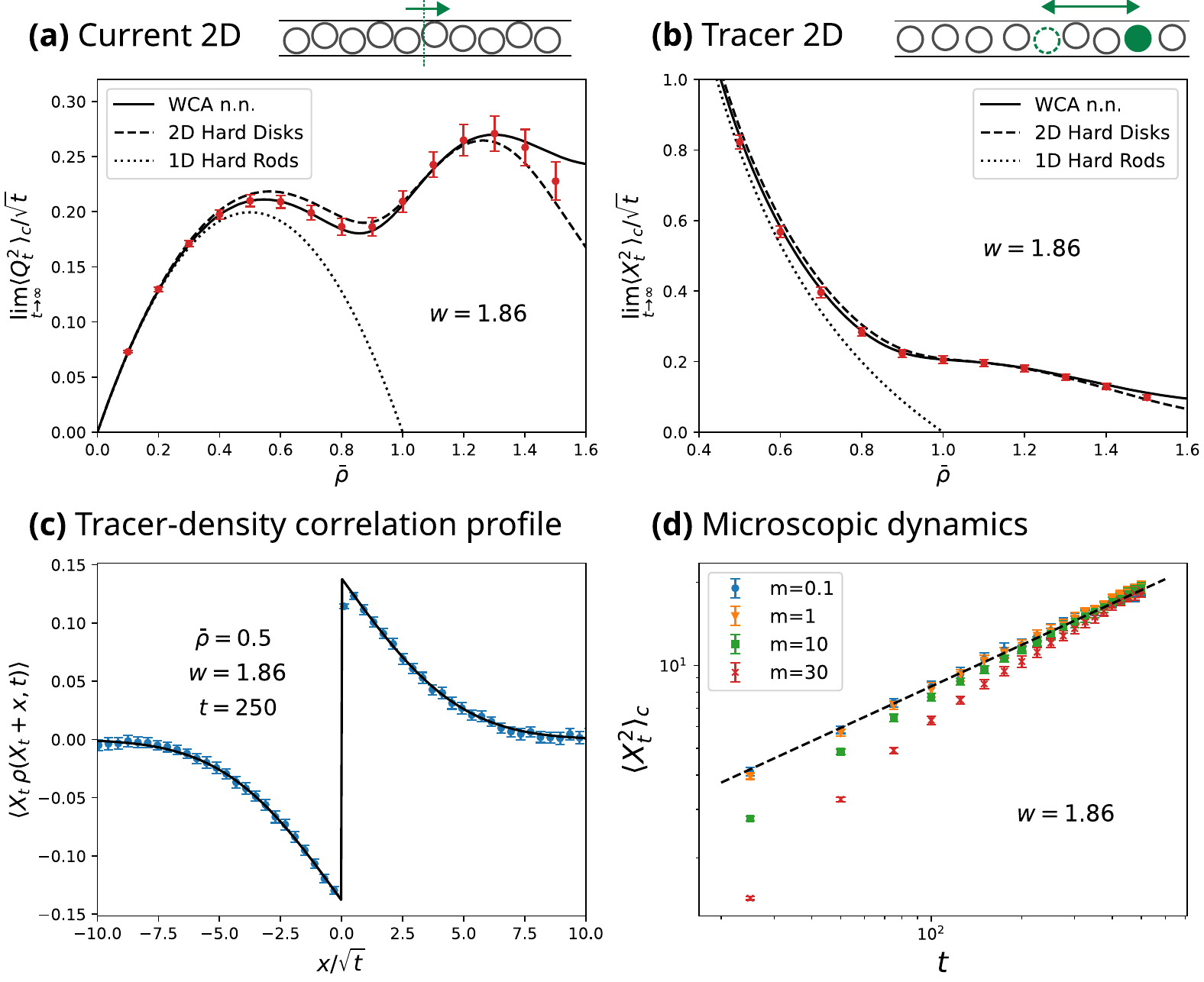}
\caption{\textbf{Robustness with respect to interaction, initial preparation and dynamics.} \textbf{(a)(b)} The effect persists for a smooth interaction and a quenched initial preparation. The points are obtained from numerical simulations performed with LAMMPS~\cite{Thompson:2022} with $N=800$ particles in $2D$ channels of different length $L$ and width $w = 1.86$ with Weeks-Chandler-Andersen potential~\cite{Weeks:1971}, damping factor $\gamma = 1/\mu_0 = 1$, temperature $T = 1$ and mass $m=0.1$. Initially, the particles are equally spaced, corresponding to a quenched preparation. The integrated current and tracer displacement are measured every $\delta t = 25$, and the prefactor is obtained from a log--log fit over the time window
$100\leq t\leq500$. These values have been selected to be beyond the crossover to the single-file regime and before the emergence of finite-size effects (see SI for more details). The results are averaged over $200$ simulations, over all the particles for the tracer and over $50$ measurement points for the current. Error bars indicate 95\% bootstrap confidence intervals obtained by resampling the $200$ realizations.
\textbf{(c)} Bath--tracer density profile associated with a tracer displacement for a quenched initial condition. The profile shows the large-scale rearrangement of the surrounding particles in front of and behind the tracer. The points are obtained from the same simulations as in \textbf{(a,b)}, at time $t=250$. The results are averaged over $200$ simulations and the error bars indicate  95\% bootstrap confidence intervals obtained by resampling these realizations. \textbf{(d)} Microscopic dynamics: underdamped inertia affects only the crossover to the long-time diffusive regime and leaves the asymptotic single-file behaviour unchanged. The simulations are performed with LAMMPS using $N=4000$ particles in a $2d$ channel of length $L=8000$ and width $w = 1.86$ with Weeks-Chandler-Andersen potential, damping factor $\gamma = 1/\mu_0 = 1$, temperature $T = 1$ and a quenched initial condition for different masses of the particles. The results are averaged over 15 simulations and over all the particles acting as a tracer. The error bars indicate  95\% bootstrap confidence intervals obtained by resampling the $15$ realizations. The dashed line is the quenched long-time theoretical prediction for the tracer, namely~\eqref{eq:tracer_variance} divided by $\sqrt{2}$.
}
\label{fig:robustnessNum}
\end{figure}

\textit{Robustness.---}
The effect is robust in several respects. First, it is robust to the interaction potential. The hard-core case gives the simplest explicit example, but the same mechanism applies to interactions that preserve the single-file order. For instance, we show in Fig.~\ref{fig:robustnessNum}(a) the amplitude of the current fluctuations obtained from numerical simulations of particles interacting pairwise with a Weeks-Chandler-Andersen potential~\cite{Weeks:1971}, used as a smooth regularization of the hard-core interaction of size $s=1$. Although this is not a purely nearest-neighbour interaction, the qualitative behaviour is the same as in the hard-core case. We are also able to accurately reproduce these numerical results over a wide range of densities by relying on the equation of state resulting from~\eqref{eq:eos_eigenproblem} within a nearest-neighbour approximation.
The same effect is observed for tracer fluctuations, as shown in  Fig.~\ref{fig:robustnessNum}(b).
Second, it is robust to the channel geometry and dimension, as illustrated in Fig.~\ref{fig:robustness}. In a circular three-dimensional channel, the dominant transverse mode is rotationally invariant and the eigenvalue problem reduces to a radial one. In a trench, gravity enters through an additional Boltzmann weight in the transverse operator; strong vertical confinement recovers the planar-channel limit, while weaker confinement can be treated by the same three-dimensional formulation as long as the particle order is preserved. Details are given in the SI. Third, it is robust to the initial preparation: annealed (Fig.~\ref{fig:robustness}) and quenched (Fig.~\ref{fig:robustnessNum}) initial conditions change the  single-file prefactors, but the finite-width dependence still enters through the same coefficients \(D_S(\rho)\) and \(\sigma(\rho)\). Finally, it is robust to the microscopic dynamics: for underdamped inertial particles, inertia only affects the crossover to the long-time diffusive regime, as shown in Fig.~\ref{fig:robustnessNum}(d). These aspects are detailed in the SI.
Thus these finite-width effects are not a peculiarity of hard particles in a planar channel, but a direct consequence of combining preserved longitudinal order with transverse equilibrium degrees of freedom.

\textit{Discussion.---}
We first clarify the status of the results obtained above. For the class of diffusive confined systems considered here, the fluctuating hydrodynamic equation is exact at leading order in the large-scale limit, under the standard local-equilibrium assumption, which includes transverse equilibration. Once the confined equation of state $P_{\mathrm{eff}}(\rho)$ is known, the expressions derived above give the exact leading long-time asymptotics of the integrated-current fluctuations and, in the no-passing regime, of the tracer fluctuations. Neither the no-passing condition nor the nearest-neighbour assumption is required for the hydrodynamic equation itself: the former is needed only to relate tracer motion to the density field, while the latter enters only in the transfer-matrix determination of $P_{\mathrm{eff}}(\rho)$. This determination is exact for nearest-neighbour systems, including hard particles in the geometrical regime specified above; for interactions such as the WCA potential, only the evaluation of the confined equation of state is approximate.

\begin{figure}[t]
\centering
\includegraphics[width=0.9\textwidth]{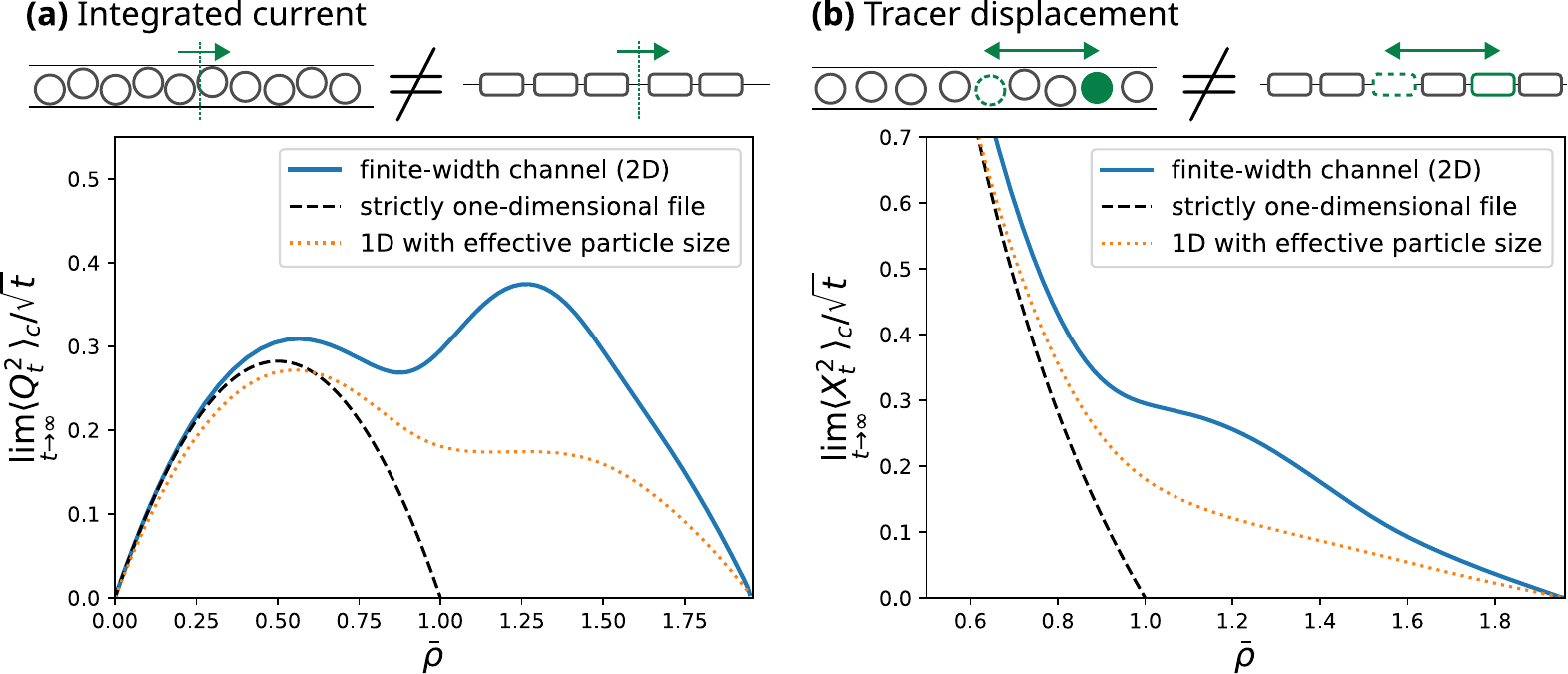}
\caption{\textbf{Comparison with effective strictly one-dimensional
descriptions.} Amplitudes of the fluctuations of \textbf{(a)} the integrated
current and \textbf{(b)} the displacement of a tracer, obtained from our exact
description of hard disks of diameter $s=1$ in a two-dimensional channel of
width $w=1.86$ (solid blue), compared with those of strictly one-dimensional
hard rods of size $s_{\mathrm{eff}}$. We show the strictly one-dimensional
system with $s_{\mathrm{eff}}=1$, used in Ref.~\cite{Lin:2005} (dashed), and
the density-dependent value of $s_{\mathrm{eff}}$ obtained by mapping the true
equation of state $P_{\mathrm{eff}}(\rho)$ onto that of hard rods, as in
Ref.~\cite{Zarif:2020} (dotted). Neither of these one-dimensional reductions
reproduces the finite-width single-file behaviour, either quantitatively or
qualitatively.
}
\label{fig:Comp1Deff}
\end{figure}

We now return to the comparison with existing approaches that describe the finite-width system using a one-dimensional hard-rod model with an effective size $s_{\mathrm{eff}}$~\cite{Lin:2005,Zarif:2020}\footnote{Except in the two limiting regimes treated analytically in
Ref.~\cite{Mon:2005}, the effective size is extracted from Monte Carlo
simulations by matching the asymptotic mobility to that of one-dimensional
hard rods; no closed analytical prescription is provided for the generic
case.}.
More precisely, Ref.~\cite{Lin:2005} uses the strictly one-dimensional size $s_{\mathrm{eff}}=s$, while Ref.~\cite{Zarif:2020} introduces a density-dependent size $s_{\mathrm{eff}}(\rho)$ by equating the confined equation of state $P_{\mathrm{eff}}(\rho)$ to that of hard rods, $P_{\mathrm{hr}}(\rho)=\rho k_{\mathrm{B}}T/(1-\rho s_{\mathrm{eff}}(\rho))$. 
We show in Fig.~\ref{fig:Comp1Deff} that these approaches are unable to reproduce the correct behaviour of the current and tracer fluctuations, either quantitatively or qualitatively.
The appropriate one-dimensional reduction is instead the fluctuating
hydrodynamics~\eqref{eq:fh}, whose constitutive coefficients
~\eqref{eq:mobility}--\eqref{eq:diffusivity_pressure} retain the full confined
equation of state rather than a single effective particle size.

These exact results are expected to have strong experimental signatures.
First, we provide the explicit range of parameters for which the effects are strong\footnote{Existing experimental measurements of tracer displacement for colloids in narrow single-file channels~\cite{Wei:2000,Lutz:2004,Lin:2005} have mostly investigated the low-density regime or channels with $w$ close to $s$, for which the finite-width effects
are small, as shown in Fig.~\ref{fig:robustness}. For instance, Ref.~\cite{Lutz:2004} explores the density range $0.3<\bar\rho s<0.6$.}: channel widths $w\geq1.7s$, while remaining within the no-passing regime
$w<2s$, and densities $\bar\rho s\geq1$. Second, the effect is both qualitative and quantitative, as shown in
Figs.~\ref{fig:effect0}--\ref{fig:Comp1Deff}. Third, the qualitative
signatures persist across interaction potentials, channel geometries,
initial preparations and microscopic dynamics, as shown in
Figs.~\ref{fig:robustness} and~\ref{fig:robustnessNum}.

\textit{Conclusion.---} For simplicity, we have focused here on the fluctuations of the observables considered above, namely the tracer displacement and the integrated current, together with their correlations with the particle density. However, the formalism presented here can give access to the full distribution of these observables, in particular through higher-order cumulants. For instance, the fourth cumulants \(\left\langle X_t^4 \right\rangle_c\) and \(\left\langle Q_t^4 \right\rangle_c\), which measure deviations from Gaussian fluctuations, have recently been computed from the fluctuating hydrodynamic equation~\eqref{eq:fh} in Ref.~\cite{Grabsch:2024b}. Therefore, combining the evolution equation~\eqref{eq:fh} with the transport coefficients~\eqref{eq:mobility} and~\eqref{eq:diffusivity_pressure} exactly determines the full large-scale fluctuation statistics of a confined system within the single-file regime.

Finite-width single-file transport is therefore not a perturbation of the strictly one-dimensional single-file problem. The coarse-grained density remains one-dimensional at large scales, but it does not remove the transverse degrees of freedom from the thermodynamics. These degrees of freedom determine the pressure of the confined ordered fluid and, through it, the collective diffusivity, current fluctuations, tracer fluctuations and density profiles. This provides a direct route from the equilibrium geometry of a channel to the non-equilibrium fluctuations of a file, and identifies quasi-one-dimensional single-file transport as a distinct regime of confined many-body dynamics.

\clearpage
%

\end{document}


\setcounter{equation}{0}
\setcounter{figure}{0}
\setcounter{table}{0}
\setcounter{page}{1}
\makeatletter
\renewcommand{\theequation}{S\arabic{equation}}
\renewcommand{\thefigure}{S\arabic{figure}}
\renewcommand{\bibnumfmt}[1]{[S#1]}
\renewcommand{\citenumfont}[1]{S#1}

\title{Supplementary Information for\texorpdfstring{\\}{} Transverse geometry reshapes current and tracer fluctuation amplitudes in quasi-one-dimensional single files}

\author{Olivier Bénichou}
\author{Aurélien Grabsch}

\maketitle

\tableofcontents

\section{Fluctuating hydrodynamics of interacting Brownian particles}
\label{sec:FlucHydro}

We consider a system of particles of size $s$ confined in a channel of section $S$. The particles have a pairwise interaction potential $V$ that encompasses hard-core replusion and optionally an additional longer range interaction term (e.g. dipolar interaction).
The positions $\V{r}_n=(x_n,\V{y}_n)$ of the particles evolve according to the set of coupled Langevin equations
\begin{equation}
	\dt{\V{r}_n}{t}
	=-\mu_0 \sum_{m \neq n} \nabla V(\V{r}_n-\V{r}_{m})
	+\sqrt{2\mu_0 k_{\rm B}T}\,\V{\eta}_n(t)
	\:,
	\label{eq:langevin_channel_all}
\end{equation}
with reflecting conditions on the boundaries of the channel. Here $\mu_0$ is the microscopic mobility and $\V{\eta}_n$ is a vector Gaussian white noise with
\begin{equation}
	\label{eq:CorrelNoise0}
	\moy{ (\V{\eta}_n)_i(t) (\V{\eta}_m)_j(t') }
	= \delta_{n,m} \: \delta_{i,j} \: \delta(t-t')
	\:.
\end{equation}
The microscopic density of particles
\begin{equation}
	\label{eq:DefMicroDens}
	\rhom(\V{r},t)
	\equiv
	\sum_n \delta( \V{r} - \V{r}_n(t) )
\end{equation}
is known to obey the Dean-Kawasaki equation~\cite{Kawasaki:1994,Dean:1996,Illien:2024}
\begin{equation}
	\label{eq:DKeq}
	\partial_t \rhom
	+ \V{\nabla} \cdot \jm
	= 0
	\:,
	\quad
	\jm 
	=-  D_0 \V{\nabla} \rhom
	- \mu_0 \rhom \V{\nabla}( V \star \rhom)
	- \sqrt{2 D_0 \rho_0} \: \V{\eta}
	\:,
\end{equation}
where we have denoted $D_0 = \mu_0 k_{\mathrm{B}} T$ the single-particle diffusion coefficient, the convolution
\begin{equation}
	\label{eq:DefConvol}
	(V \star \rhom)(\V{r},t)
	\equiv
	\int V(\V{r}-\V{r}') \rhom(\V{r}',t) \dd^d \V{r}' 
\end{equation}
encodes the pairwise interaction between the particles and $\V{\eta}$ is a Gaussian white noise in space and time,
\begin{equation}
	\label{eq:NoiseCorrelDK}
	\moy{\eta_i(\V{r},t) \eta_j(\V{r}',t')}
	= \delta_{i,j} \delta(t-t') \delta^{(d)}(\V{r}-\V{r}')
	\:
\end{equation}

\bigskip

From the Dean-Kawasaki equation~\eqref{eq:DKeq}, we derive the effective one-dimensional fluctuating hydrodynamics equation~(3). For this, we define the coarse-grained density and current by averaging the microscopic density $\rhom$ and current $\jm$ on a mesoscopic length $\ell$ and across the section $S$ of the channel
\begin{equation}
	\label{eq:DefMacroDens}
	\rho(x,t) = \frac{1}{\ell} \int_0^\ell \dd x' \int_S \dd^{d-1} \V{y} \:
	\rhom \big( (\Lambda x + x', \V{y}), \Lambda^2 t \big)
	\:,
\end{equation}
\begin{equation}
	\label{eq:DefMacroCurr}
	j(x,t) = \frac{\Lambda}{\ell} \int_0^\ell \dd x' \int_S \dd^{d-1} \V{y} \:
	\jm \big( (\Lambda x + x', \V{y}), \Lambda^2 t \big) \cdot \V{e}_x
	\:,
\end{equation}
where $\Lambda \gg 1$ is the ratio between the macroscopic and the microscopic scales. Note that we have performed a diffusive rescaling (space by $\Lambda$ and time by $\Lambda^2$) since we expect that the system remains diffusive at large scales. As we will see later, the precise choice of $\Lambda$ is irrelevant, as long as $\Lambda \gg 1$.

To derive the equations satisfied by the macroscopic fields~(\ref{eq:DefMacroDens},\ref{eq:DefMacroCurr}), we adapt the approach used in the unconfined geometry~\cite{Das:2013} (see also the recent work~\cite{Saha:2026}).

First, we compute
\begin{align}
	\partial_x j
	&=
	\frac{\Lambda^2}{\ell} \int_0^\ell \dd x' \int_S \dd^{d-1} \V{y} \:
	\partial_{x'}\jm \big( (\Lambda x + x', \V{y}), \Lambda^2 t \big) \cdot \V{e}_x
	\nonumber
	\\
	&=
	- \frac{1}{\ell} \int_0^\ell \dd x' \int_S \dd^{d-1} \V{y} \:
	\partial_t \rhom \big( (\Lambda x + x', \V{y}), \Lambda^2 t \big)
	- \frac{\Lambda^2}{\ell} \int_0^\ell \dd x' \int_S \dd^{d-1} \V{y} \:
	\V{\nabla}_{\V{y}} \cdot \jm \big( (\Lambda x + x', \V{y}), t \big)
	 \:,
\end{align}
where we have used the continuity equation~\eqref{eq:DKeq} and denoted $\V{\nabla}_{\V{y}} \cdot = (\V{\nabla} \cdot) - (\partial_x \V{e}_x \cdot)$ the divergence of the $\V{y}$ components only. Since particles cannot escape from the boundaries of the channel, the normal component of $\jm$ vanishes on the boundaries, hence the second term is zero. We thus obtain that the one-dimensional macroscopic fields obey the continuity equation
\begin{equation}
	\label{eq:ContMacro}
	\partial_t \rho + \partial_x j = 0
	\:,
\end{equation}
as expected. 

Second, inserting the expression of the microscopic current~\eqref{eq:DKeq} into the definition of the macroscopic current~\eqref{eq:DefMacroCurr}, we can write
\begin{equation}
	\label{eq:MacroCurrCG}
	j(x,t) = j_{\mathrm{det}}(x,t)
	+ j_{\mathrm{stoch}}(x,t)
	\:,
\end{equation}
where
\begin{multline}
	\label{eq:DefJdet}
	j_{\mathrm{det}}(x,t) \equiv - D_0 \partial_x \rho(x,t)
	\\
	- \frac{\Lambda}{\ell} \mu_0 \int_0^\ell \dd x' \int_S \dd^{d-1} \V{y} \:
	\rhom((\Lambda x + x', \V{y}), \Lambda^2 t)
	\int_{-\infty}^\infty \dd z \int_S \dd \V{y}' \:
	\partial_{x'} V( (\Lambda x + x' - z, \V{y} - \V{y}') ) \rhom( (z,\V{y'}), \Lambda^2 t)
	\:,
\end{multline} 
\begin{equation}
	\label{eq:DefJstoch}
	j_{\mathrm{stoch}}(x,t)
	\equiv
	- \frac{\Lambda}{\ell} \int_0^\ell \dd x' \int_S \dd^{d-1} \V{y} \:
	\sqrt{2 D_0 \rhom\big( (\Lambda x + x', \V{y}), \Lambda^2 t \big)}
	\: \eta_x \big( (\Lambda x + x', \V{y}), \Lambda^2 t \big)
	\:.
\end{equation}
By making a change of variables, we can write the deterministic part~\eqref{eq:DefJdet} as
\begin{multline}
	\label{eq:Jdet2}
	j_{\mathrm{det}}(x,t) = - D_0 \partial_x \rho(x,t)
	\\
	- \frac{\Lambda}{\ell} \mu_0 
	\int_{-\infty}^\infty \dd z \int_S \dd \V{y} \dd \V{y}' \:
	\partial_{z} V( (z, \V{y} - \V{y}') )
	\int_0^\ell \dd x'
	 \:
	\rhom((\Lambda x + x', \V{y}), \Lambda^2 t)
	 \rhom( (\Lambda x + x' - z,\V{y'}), \Lambda^2 t)
	\:,
\end{multline} 
For a sufficiently large mesoscopic cell $\ell$, the integral over $x'$ can be interpreted as an equilibrium average in an infinite system with mean density $\rho(x,t)$ (since the macroscopic density is constant on the mesoscopic fluid cell),
\begin{equation}
	\label{eq:LocalEq}
	\frac{1}{\ell} \int_0^\ell \dd x'
	 \:
	\rhom((\Lambda x + x', \V{y}), \Lambda^2 t)
	 \rhom( (\Lambda x + x' - z,\V{y'}), \Lambda^2 t)
	 = \moy{
	 	\rhom((0, \V{y}) )
	 	\rhom( (- z,\V{y'}) )
	 }_{\text{l.e. } \rho(x,t)}
	 \:,
\end{equation}
where we have denoted $\moy{\cdots}_{\text{l.e. } \rho(x,t)}$ the averaging with the local equilibrium measure, at density $\rho(x,t)$. However, naively performing this substitution into the expression of the current~\eqref{eq:Jdet2} yields $0$ since the two point function is symmetric in $z$ and $\partial_z V$ is anti-symmetric. One should thus in principle look for corrections to the local equilibrium, which are in practice intractable. Alternatively, we can first rewrite the current in~\eqref{eq:Jdet2} in a more convenient form, involving the Irving-Kirkwood tensor~\cite{Irving:1950} before using the local equilibrium property (see also Ref.~\cite{Saha:2026} for a related approach).
Indeed, using that $\partial_z V$ is anti-symmetric, we replace it by $\frac{1}{2}(\partial_z V(z) - \partial_z V(-z))$. Making the change of variables $z \to -z$ in the second term and regrouping them, we obtain
\begin{multline}
	\label{eq:Jdet3}
	j_{\mathrm{det}}(x,t) = - D_0 \partial_x \rho(x,t)
	+ \frac{\Lambda}{2 \ell} \mu_0 
	\int_{-\infty}^\infty \dd z \int_S \dd \V{y} \dd \V{y}' \:
	\partial_{z} V( (z, \V{y} - \V{y}') )
	\\
	\times
	\int_0^\ell \dd x' \int_0^1 \dd u 
	 \: \dt{}{u} \Big[
	\rhom((\Lambda x + x' + u z, \V{y}), \Lambda^2 t)
	 \rhom( (\Lambda x + x' - (1-u)z,\V{y'}), \Lambda^2 t)
	 \Big]
	\:.
\end{multline}
Equivalently,
\begin{multline}
	\label{eq:Jdet4}
	j_{\mathrm{det}}(x,t) = - D_0 \partial_x \rho(x,t)
	+ \frac{1}{2 \ell} \mu_0  \partial_x
	\int_{-\infty}^\infty \dd z \int_S \dd \V{y} \dd \V{y}' \:
	z \partial_{z} V( (z, \V{y} - \V{y}') )
	\\
	\times
	\int_0^\ell \dd x' \int_0^1 \dd u 
	 \:
	\rhom((\Lambda x + x' + u z, \V{y}), \Lambda^2 t)
	 \rhom( (\Lambda x + x' - (1-u)z,\V{y'}), \Lambda^2 t)
	\:.
\end{multline}
Using now the local equilibrium property~\eqref{eq:LocalEq}, we obtain a non-trivial result
\begin{equation}
	\label{eq:Jdet5}
	j_{\mathrm{det}}(x,t) = - \mu_0 \partial_x \left[  
	k_{\mathrm{B}} T \rho(x,t)
	- \frac{\mu_0}{2}
	\int_{-\infty}^\infty \dd z \int_S \dd \V{y} \dd \V{y}' \:
	z \partial_{z} V( (z, \V{y} - \V{y}') )
	\moy{
	 	\rhom((0, \V{y}) )
	 	\rhom( (z,\V{y'}) )
	 }_{\text{l.e. } \rho(x,t)}
	 \right]
	\:.
\end{equation}
In the bracket, we recognize the virial equation for the equilibrium longitudinal pressure $P_{\mathrm{long}}(\rho)$ in a channel~\cite{Irving:1950},
\begin{equation}
	\label{eq:JdetPress}
	j_{\mathrm{det}}(x,t) = - \mu_0 S \: \partial_x P_{\mathrm{long}}(\rho(x,t))
	= - \mu_0 \: \partial_x P_{\mathrm{eff}}(\rho(x,t))
	\:,
\end{equation}
where $P_{\mathrm{eff}}(\rho) = S P_{\mathrm{long}}(\rho)$ is the effective one dimensional pressure.

For a given microscopic density $\rhom$, the stochastic term~\eqref{eq:DefJstoch} is Gaussian, with $\moy{j_{\mathrm{stoch}}(x,t)} = 0$. Hence it is fully determined by its two-point correlation function
\begin{multline}
	\moy{
		j_{\mathrm{stoch}}(x,t)
		j_{\mathrm{stoch}}(x',t')
	}
	= 
	\frac{2 D_0 \Lambda^2}{\ell^2} 
	\int_0^\ell \dd z \dd z' \int_S \dd \V{y} \dd \V{y}' \:
	\sqrt{\rhom\big( (\Lambda x + z, \V{y}), \Lambda^2 t \big)
	\rhom\big( (\Lambda x' + z', \V{y}'), \Lambda^2 t' \big)}
	\\
	\times
	\moy{ \eta_x \big( (\Lambda x + z, \V{y}), \Lambda^2 t \big)
	\eta_x \big( (\Lambda x' + z', \V{y}'), \Lambda^2 t' \big)
	}
	\:.
\end{multline}
Using the noise correlator~\eqref{eq:NoiseCorrelDK}, this reduces to
\begin{equation}
	\moy{
		j_{\mathrm{stoch}}(x,t)
		j_{\mathrm{stoch}}(x',t')
	}
	= 
	\frac{2 D_0 \Lambda^2}{\ell^2} 
	\int_0^\ell \dd z \int_S \dd \V{y} \:
	\rhom\big( (\Lambda x + z, \V{y}), \Lambda^2 t \big)
	\delta \big(\Lambda^2 (t-t') \big)
	\mathbb{1}_{[0,\ell]} \big(\Lambda(x-x') + z \big)
	\:,
\end{equation}
with $\mathbb{1}_{[0,\ell]}(x) = 1$ if $x \in [0,\ell]$ and $0$ otherwise. Furthermore, since $\int \mathbb{1}_{[0,\ell]}(\Lambda x) \dd x = \ell \Lambda^{-1}$, and $\Lambda \gg 1$, we obtain,
\begin{align}
	\moy{
		j_{\mathrm{stoch}}(x,t)
		j_{\mathrm{stoch}}(x',t')
	}
	&= 
	\frac{2 D_0}{\Lambda \ell} 
	\int_0^\ell \dd z \int_S \dd \V{y} \:
	\rhom\big( (\Lambda x + z, \V{y}), \Lambda^2 t \big)
	\delta(x-x') \delta(t-t')
	\nonumber
	\\
	&= \frac{2 D_0 \rho(x,t)}{\Lambda} \delta(x-x') \delta(t-t')
	\:.
\end{align}
Therefore, we can write that
\begin{equation}
	\label{eq:JStochFin}
	j_{\mathrm{stoch}} = - \sqrt{\frac{2D_0 \rho}{\Lambda}} \: \eta
	\:,
\end{equation}
where $\eta$ is now a $(1+1)$-dimensional Gaussian white noise, with
\begin{equation}
	\label{eq:NoiseCorrel1D}
	\moy{\eta(x,t) \eta(x',t')} = \delta(x-x') \delta(t-t')
	\:.
\end{equation}

Finally, combining the expressions of the deterministic~\eqref{eq:JdetPress} and stochastic~\eqref{eq:JStochFin} of the macroscopic current~\eqref{eq:MacroCurrCG} and inserting them into the continuity equation~\eqref{eq:ContMacro}, we obtain the fluctuating hydrodynamics equation~(3),
\begin{equation}
	\label{eq:FlucHydroSM}
	\boxed{
	\partial_t \rho
	= \partial_x \left[
		D_S(\rho) \partial_x \rho
		+ \sqrt{\frac{\sigma(\rho)}{\Lambda}} \: \eta
	\right]	
	\:,
	}
\end{equation}
with the two transport coefficients
\begin{equation}
	\label{eq:TrCoefsSM}
	\boxed{
	D_S(\rho) = \mu_0 P_{\mathrm{eff}}'(\rho)
	\:,
	\quad
	\sigma(\rho) = 2 D_0 \rho
	\:.
	}
\end{equation}

\bigskip

We can check that these transport coefficients indeed satisfy the fluctuation-dissipation relation~\cite{Bertini:2015,Derrida:2025a}
\begin{equation}
	\label{eq:FlucDiss}
	\frac{2 k_{\mathrm{B}}T D_S(\rho)}{\sigma(\rho)}
	= f_S''(\rho)
	\:,
\end{equation}
$f_S = F/L$ is the free energy $F$ per unit length for a system of length $L$ and section $S$. Indeed, denoting $N = \rho L$ the number of particle and using that
\begin{equation}
	\label{eq:DefEffPress}
	P_{\mathrm{eff}}(\rho) = 
	- \left( \dep{F}{L} \right)_{T,N}
	= \rho f_S'(\rho) - f_S(\rho)
	\:,
\end{equation}
we deduce
\begin{equation}
	P_{\mathrm{eff}}'(\rho) = \rho f_S''(\rho)
	\:.
\end{equation}
Combining the expressions of $D_S$ and $\sigma$~\eqref{eq:TrCoefsSM} with this last relation, we recover the fluctuation-dissipation relation~\eqref{eq:FlucDiss}, as it should.

\section{Equilibrium pressure for a single-file system}
\label{sec:EqPr}

The use of the fluctuating hydrodynamics equation~\eqref{eq:FlucHydroSM} requires the determination of the equilibrium effective pressure~\eqref{eq:DefEffPress} to determine the transport coefficients~\eqref{eq:FlucDiss}. This is the goal of this Section.

\subsection{Derivation}

Let us consider a system of $N+1$ particles in a single file geometry. We assume that the particles interact only with their nearest neighbours. Comparing to the Langevin equations~\eqref{eq:langevin_channel_all}, this is exact for interactions located at the surface of the particles, like hard-core repulsion or for sticky hard spheres~\cite{Baxter:1968,Percus:1982}. For longer range interaction, this is an approximation valid at low density or weak interaction.
The nearest-neighbour interaction together with the single-file constraint makes the thermodynamic properties of this system computable exactly~\cite{Percus:1982}. We present here a similar derivation.

We consider that the particles are on the positive axis $x>0$, and denote $\V{r}=(x, \V{y})$ the position of the last particle. We denote $Z_{N}(\beta; x, \V{y})$ the partition function of the remaining $N$ particles. Since the particles are always ordered, we can express this partition function as
\begin{equation}
  \label{eq:PartFct0}
  Z_N(\beta; x, \V{y})
  = \frac{1}{\ell_0^{Nd}} \int_0^x \dd x_1 \int_S \dd^{d-1} \V{y}_1
  \e^{-\beta V(\V{r} - \V{r}_1)}
  \cdots \int_0^{x_{N-1}} \dd x_{N}\int_S \dd^{d-1} \V{y}_N
  \: \e^{-\beta V(\V{r}_{N-1} - \V{r}_N)}
  \:,
\end{equation}
where we have introduced $\ell_0 = \sqrt{\beta h^2/(2\pi m)}$ the De Broglie
wavelength, which comes from the integration over the momenta. This constant is present for the normalisation but does not play a role in the following, as we will see.
 We introduce the grand-canonical version
\begin{equation}
  \label{eq:XiInvOp}
  \Xi(\beta,\mu; x, \V{y})
  = \sum_{N=0}^\infty
  \e^{\beta \mu N} Z_N(\beta; x, \V{y})
  = ( 1 - \mathscr{L})^{-1}[1](x, \V{y})
  \:,
\end{equation}
with
\begin{equation}
  \label{eq:DefIntegOp}
  \mathscr{L}[f](x, \V{y})
  \equiv \frac{\e^{\beta \mu}}{\ell_0^d} \int_0^x \dd x' \int_S \dd^{d-1} \V{y}'
  \e^{-\beta V(\V{r} - \V{r}')} f(x',\V{y}')
  \:.
\end{equation}
Hence, we find that $\Xi$ obeys the integral equation
\begin{equation}
  \label{eq:IntegEqXi}
  \Xi(\beta,\mu; x, \V{y})
  - \frac{\e^{\beta \mu}}{\ell_0^d} \int_0^x \dd x' \int_S \dd^{d-1} \V{y}'
  \e^{-\beta V(\V{r} - \V{r}')} \Xi(\beta, \mu; x',\V{y}')
  = 1
  \:.
\end{equation}
The spatial convolution can be handled by taking a Laplace
transform. We thus introduce
\begin{equation}
  \label{eq:DefLaplaceXi}
  \hat\Xi(\beta,\mu; u, \V{y}) \equiv
  \int_0^\infty \e^{-u x} \Xi(\beta,\mu; x, \V{y}) \dd x
  \:.
\end{equation}
It obeys
\begin{equation}
  \label{eq:IntegEqXiLaplace}
  \hat\Xi(\beta,\mu; u, \V{y})
  = \frac{\e^{\beta \mu}}{\ell_0^d}
  \int_S \dd^{d-1} \V{y}'
  \hat\Xi(\beta,\mu; u, \V{y}') \hat{V}(\beta; u, \V{y} - \V{y'})
  + \frac{1}{u}
  \:,
\end{equation}
where we have denoted
\begin{equation}
  \label{eq:DefVLaplace}
  \boxed{
    \hat{V}(\beta; u, \V{y}) = \int_0^\infty \dd x
    \: \e^{-u x} \e^{- \beta V \left( \sqrt{x^2 + \V{y}^2} \right)}
    \:.
  }
\end{equation}
Since the grand potential $J = - P V = - P_{\mathrm{eff}} x$ for a
system of length $x$ is obtained from
\begin{equation}
  J = - k_{\mathrm{B}} T \ln \Xi(\beta,\mu; x, \V{y})
  \:,
\end{equation}
we have that
\begin{equation}
  \Xi(\beta,\mu; x, \V{y})
  \underset{x \to \infty}{\simeq} \e^{\beta P_{\mathrm{eff}} x} \phi(\beta,\mu; \V{y})
  \:.
\end{equation}
Thus taking the Laplace transform~(\ref{eq:DefLaplaceXi}) yields
\begin{equation}
  \hat\Xi(\beta,\mu; u, \V{y})
  \underset{u \to \beta P_{\mathrm{eff}}}{\simeq}
  \frac{\phi(\beta,\mu; u, \V{y})}{u - \beta P_{\mathrm{eff}}}
  \:,
\end{equation}
with $\beta P_{\mathrm{eff}}$ the rightmost pole of $\hat\Xi$.  This
gives into the integral equation~(\ref{eq:IntegEqXiLaplace})
\begin{equation}
  \label{eq:EVeqPhiSM}
  \boxed{
    \phi(\beta,\mu; \V{y}) = \frac{\e^{\beta \mu}}{\ell_0^d}
    \int_S \dd^{d-1} \V{y}' \hat{V}(\beta; \beta P_{\mathrm{eff}}, \V{y} - \V{y}')  \phi(\beta,\mu; \V{y}')
    \:.
  }
\end{equation}
This equation shows that $\phi$ is an eigenfunction of $\hat{V}$, with
eigenvalue $\frac{\e^{\beta \mu}}{\ell_0^d}$. This gives the chemical potential $\mu$ as a function of $P_{\mathrm{eff}}$ and $\beta$. Inverting this relation gives $P_{\mathrm{eff}}(\beta,\mu)$. To express the pressure in terms of the linear density $\rho$, we perform a Legendre transform to the canonical ensemble, which expresses the free energy as
\begin{equation}
  F = \min_\mu [J + \mu N ] = x  \min_\mu[ \mu \rho - P_{\mathrm{eff}}(\beta,\mu)  ]
  \:.
\end{equation}
Thus, we obtain the chemical potential $\mu(\beta,\rho)$ as the solution of
\begin{equation}
  \label{eq:LegTrForMuRhoSM}
  \boxed{
    \partial_\mu P_{\mathrm{eff}}(\beta,\mu) \Big|_{\mu(\beta,\rho)} = \rho
    \:.
  }
\end{equation}
The equation of state is then obtained as
$P_{\mathrm{eff}}(\beta, \rho) = P_{\mathrm{eff}}(\beta,\mu(\beta,\rho))$.

\subsection{Numerical evaluation}
\label{sec:NumEval}

The eigenvalue equation~\eqref{eq:EVeqPhiSM} together with the Legendre transform~\eqref{eq:LegTrForMuRhoSM} fully determine the effective pressure $P_{\mathrm{eff}}$, and thus the diffusion coefficient~\eqref{eq:TrCoefsSM}. These however cannot be solved analytically, but can be evaluated numerically. We present the numerical method used to compute this equation of state.
For simplicity, we discuss the case of a two-dimensional channel, so that $\V{y}$ is a scalar. The same discussion can be extended to arbitrary dimension.

The first step is to discretise the eigenvalue equation~\eqref{eq:EVeqPhiSM}. This can be done with equidistant points, but it is much more efficient to use a Gaussian quadrature, as in Ref.~\cite{Bornemann:2010}. The discretised equations are then
\begin{equation}
  \label{eq:DiscrEVeq}
  \sum_j \hat{V}(\beta, \beta P_{\mathrm{eff}}, y_i - y_j) w_j
  \phi_j(\beta,\mu)
  = \ell_0^d \e^{-\beta \mu} \phi_i(\beta,\mu)
  \:,
\end{equation}
with positions $y_i$ and weights $w_i$. 
We diagonalise numerically the matrix with entries
$\hat{V}(\beta, \beta P_{\mathrm{eff}}, y_i - y_j) w_j$, whose largest
eigenvalue gives $\ell_0^d \e^{-\beta \mu(P_{\mathrm{eff}},\beta)}$, and the
associated eigenvector we denote $u_i$, with the normalisation condition
\begin{equation}
  \sum_i w_i u_i^2 = 1
  \:.
\end{equation}
Note that the matrix can be made symmetric by multiplying by
$\sqrt{w}$ on the left and $1/\sqrt{w}$ on the right. Applying
standard perturbation theory to this symmetric matrix, we can express the derivative of $\mu$ with respect to $P_{\mathrm{eff}}$ in closed form,
\begin{equation}
  \dt{}{P_{\mathrm{eff}}} \left( \ell_0^d \e^{-\beta \mu(P_{\mathrm{eff}})} \right)
  = \sum_{i,j} w_i w_j \dt{}{P_{\mathrm{eff}}} \hat{V}(\beta, \beta P_{\mathrm{eff}}, y_i - y_j)
  u_i u_j
  \:.
\end{equation}
Hence
\begin{equation}
  \dt{\mu}{P_{\mathrm{eff}}}
  = - \frac{\ell_0^{-d}}{\beta}
  \e^{\beta \mu(P_{\mathrm{eff}})} \sum_{i,j} w_i w_j \dt{}{P_{\mathrm{eff}}} \hat{V}(\beta, \beta P_{\mathrm{eff}}, y_i - y_j)
  u_i u_j
  \:.
\end{equation}
Together with~(\ref{eq:LegTrForMuRhoSM}), this gives a parametric
representation of $P_{\mathrm{eff}}(\rho)$.

We increase the number of discretisation points until we reach a convergence of the eigenvalues and eigenvectors. Typically $\sim 15$ points are sufficient to reach convergence of a large domain of density $\rho$. Note that for linearly spaced points, the number of required points would be about $10$ times larger to reach the same accuracy.

\subsection{Applications}

\subsubsection{Two-dimensional channel}

\begin{figure}
    \centering
    \includegraphics[width=0.4\textwidth]{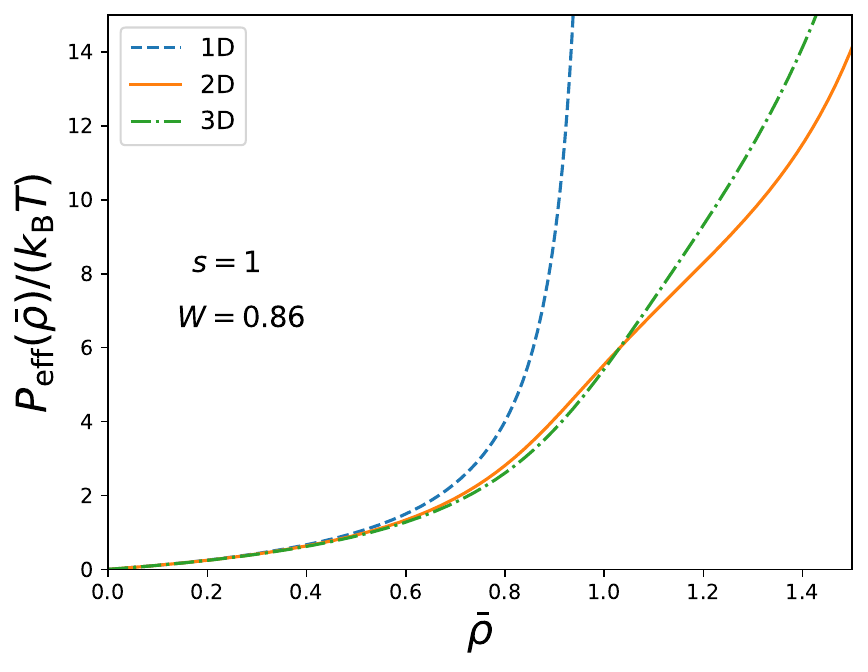}
    \includegraphics[width=0.4\textwidth]{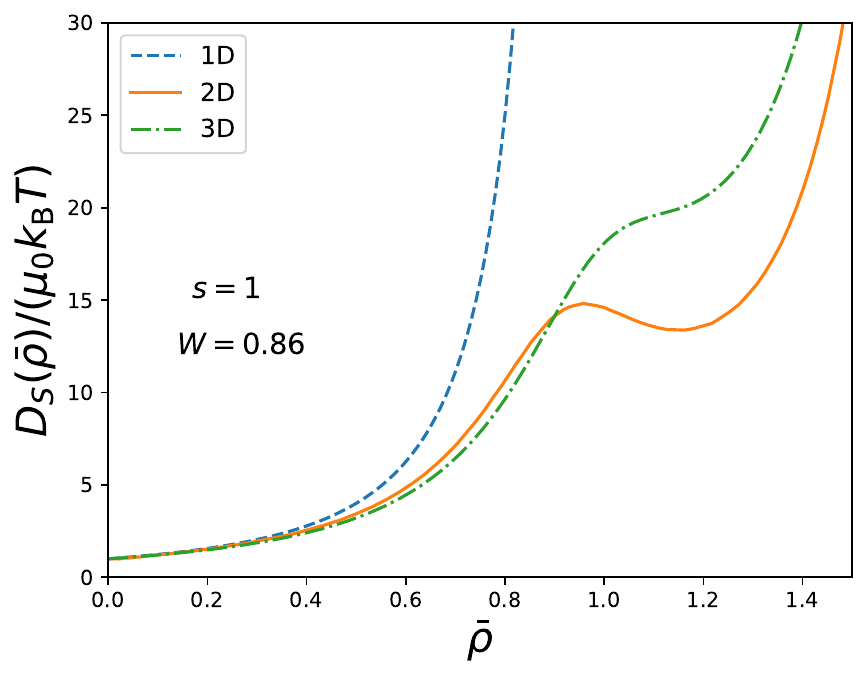}
    \caption{Left: Effective pressure $P_{\mathrm{eff}}(\rho)$ in different dimensions $d$ for hard spheres of size $s=1$ in a circular channel of width $s+W$ with $W=0.86$. Right: same for the collective diffusion coefficient $D_S(\rho)$ obtained from~\eqref{eq:TrCoefsSM}.}
    \label{fig:PandD}
\end{figure}

The case of a two-dimensional channel can be solved numerically using the procedure described in Section~\ref{sec:NumEval}. The result is shown in Fig.~\ref{fig:PandD} for hard disks.

Note that for hard disks, the condition that particles interact only with their nearest-neighbors is enforced for a channel of width $s+W$, with $W < s \sqrt{3}/2 \simeq 0.866 s$.

\subsubsection{Three-dimensional circular channel}

\begin{figure}
  \centering
  \includegraphics[width=0.3\textwidth]{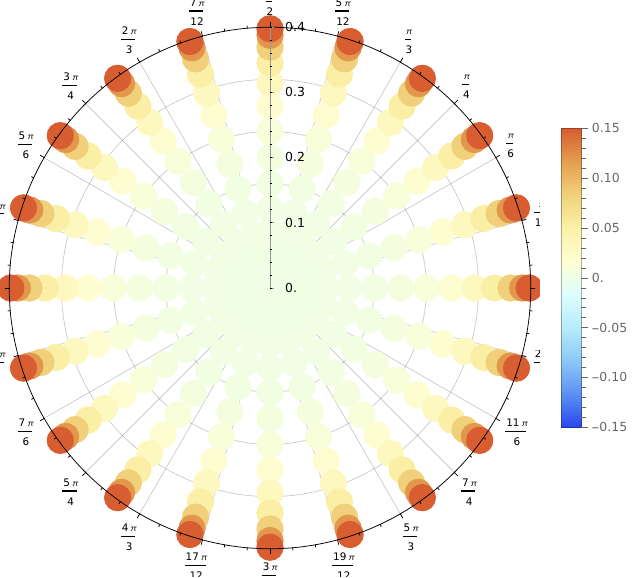}
  \includegraphics[width=0.3\textwidth]{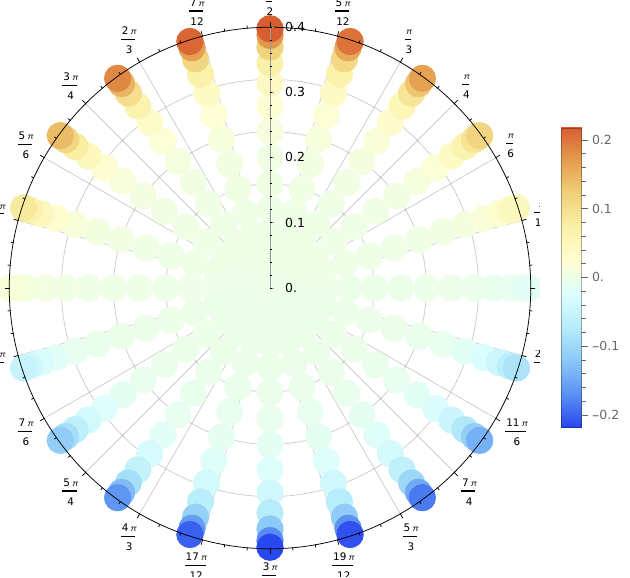}
  \includegraphics[width=0.3\textwidth]{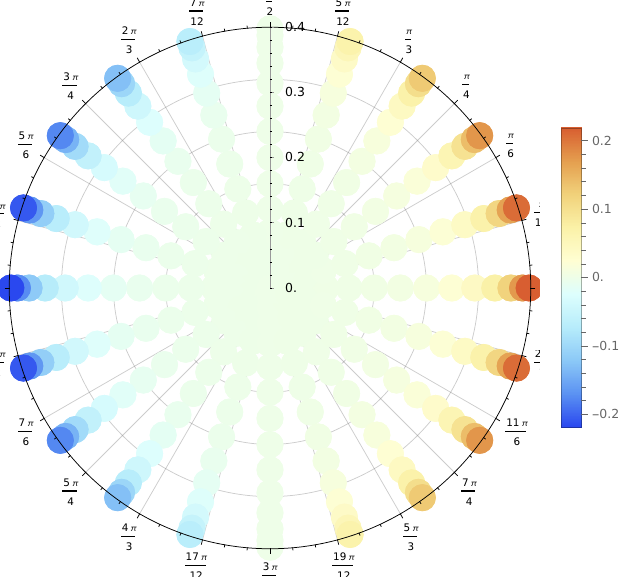}
  \caption{First eigenmodes $\phi$ of~(\ref{eq:EVeqPhiSM}) for hard
    spheres of diameter $s = 1$ in circular channel of width $s+w$
    with $w = 0.86$, with a polar discretisation of $N_r = 15$ points
    in the radial direction and $N_\theta = 20$ points in the circular
    direction. The first eigenmode (left) is invariant under rotation,
    while the next two (center and right) are not. They are computed at a pressure
    $\beta P_{\mathrm{eff}} = 20$, for $\beta = 1$.}
  \label{fig:EVcircular}
\end{figure}

The discretisation~(\ref{eq:DiscrEVeq}) can be performed in any
dimension, but this increases the dimension of the matrix to study
very quickly. In the case of a circular channel, we expect the
eigenmode associated to the largest eigenvalue to be invariant under
rotation. We have checked this fact numerically, as shown in
Fig.~\ref{fig:EVcircular}.

Thus, looking for an eigenmode of the form $\phi(\V{y}) = \phi(r)$
with $r$ the radial coordinate, the eigenvalue
equation~(\ref{eq:EVeqPhiSM}) becomes
\begin{equation}
  \label{eq:RadialEvEq}
  \boxed{
    \ell_0^d \e^{- \beta \mu} \: \phi(\beta,\mu,r)
    = \int_0^{\frac{w}{2}} \dd r' \: \phi(\beta,\mu,r') \: r' \int_{0}^{2\pi} \dd \theta \:
    \hat{V}(\beta,\beta P_{\mathrm{eff}}, \sqrt{r^2 + (r')^2 - 2 r r' \cos \theta})
    \:.
  }
\end{equation}
For any interaction potential, solving this eigenvalue equation with
the method of Section~\ref{sec:NumEval} is much more efficient than
the original $2d$ equation.

Note that for hard spheres, the condition that particles interact only with their nearest-neighbors is enforced for a circular channel of diameter $s+W$, with $W < s \sqrt{3}/2 \simeq 0.866 s$.

\subsubsection{Trenches}

In several experiments, the particles are placed in trenches in which
they are confined by gravity~\cite{Wei:2000}. We can also describe this situation, assuming that particles remain in the same order.

The partition function~(\ref{eq:PartFct0}) becomes
\begin{multline}
  \label{eq:PartFctTrench}
  Z_N(\beta; x, y, z)
  = \frac{1}{\ell_0^{3N}} \e^{- \beta m g z} \int_0^x \dd x_1 \int_0^W \dd y_1 \int_0^\infty \dd z_1 \:
  \e^{-\beta V(\V{r} - \V{r}_1)} \e^{-\beta m g z_1}
  \\
  \cdots \int_0^{x_{N-1}} \dd x_{N}  \int_0^W \dd y_N \int_0^\infty \dd z_N
  \: \e^{-\beta V(\V{r}_{N-1} - \V{r}_N)} \e^{-\beta m g z_N}
  \:,
\end{multline}
with $m$ the mass of a particle, and $g$ the gravitational field. As
before, we have
\begin{equation}
  \label{eq:XiInvOpTr}
  \Xi(\beta,\mu; x, y, z)
  = \sum_{N=0}^\infty
  \e^{\beta \mu N} Z_N(\beta; x, y, z)
  = ( 1 - \mathscr{L})^{-1}[ (x,y,z) \mapsto  \e^{-\beta m g z}](x, y, z)
  \:,
\end{equation}
with
\begin{equation}
  \label{eq:DefIntegOpTr}
  \mathscr{L}[f](x, y, z)
  \equiv \frac{\e^{\beta \mu}}{\ell_0^3} \e^{- \beta m g z} \int_0^x \dd x' \int_0^W \dd y' \int_0^\infty \dd z'
  \e^{-\beta V(\V{r} - \V{r}')} f(x',y', z')
  \:.
\end{equation}
Repeating the same steps as before, we get that the effective pressure
$P_{\mathrm{eff}}$ can be obtained as the solution of the eigenvalue
equation
\begin{equation}
  \label{eq:EvTrench}
  \boxed{
    \ell_0^3 \e^{-\beta \mu} \phi(\beta,\mu,y,z)
    = \e^{-\beta m g z} \int_0^W \dd y' \int_0^\infty \dd z'
    \hat{V}(\beta, \beta P_{\mathrm{eff}}, y-y', z-z')
    \phi(\beta,\mu,y',z')
    \:,
  }
\end{equation}
where we denoted
\begin{equation}
  \hat{V}(\beta,u,y,z)
  = \int_0^\infty \dd x \: \e^{-u x} \e^{-\beta V(\sqrt{x^2 + y^2 + z^2})}
  \:.
\end{equation}

\begin{figure}
  \centering
  \includegraphics[width=0.3\textwidth]{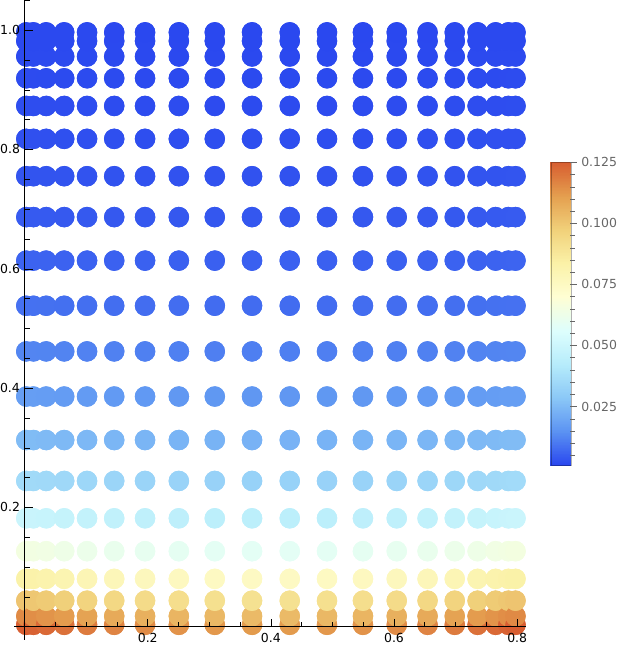}
  \includegraphics[width=0.3\textwidth]{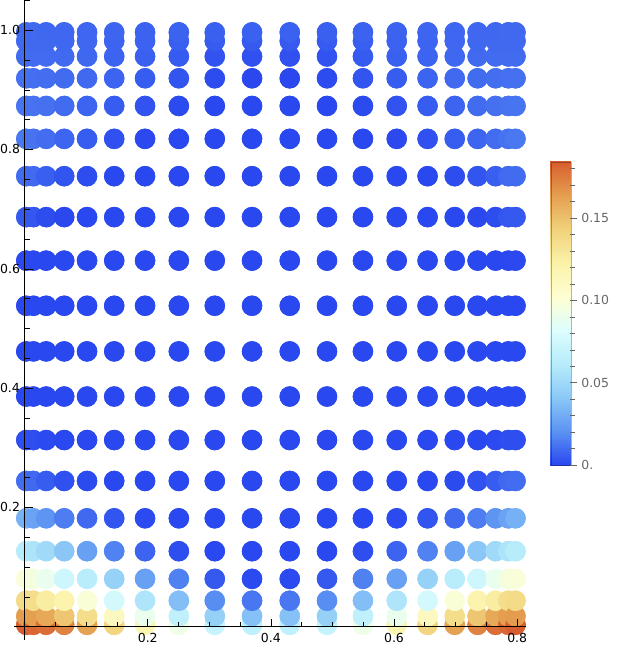}
  \includegraphics[width=0.3\textwidth]{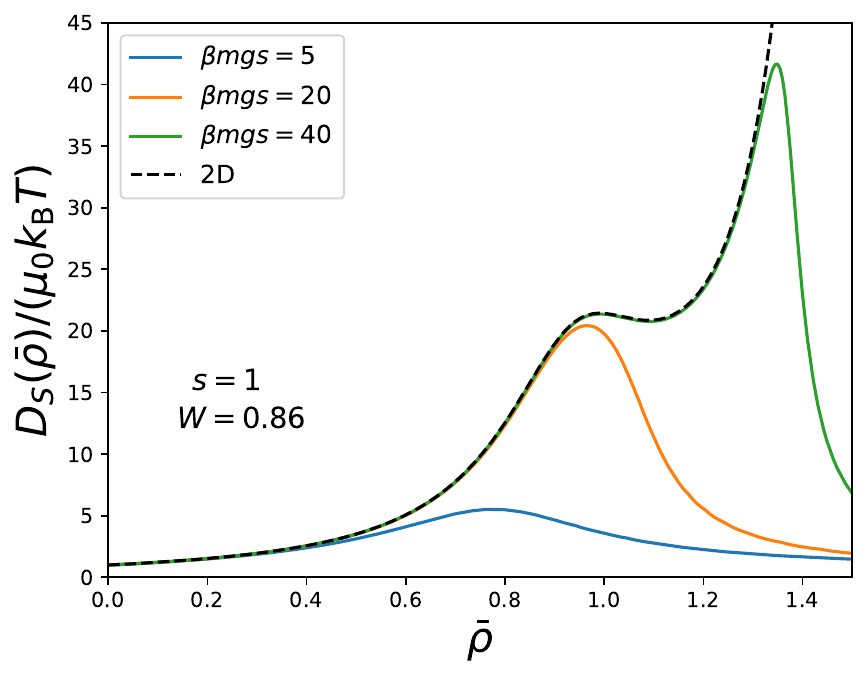}
  \caption{Eigenvector obtained from~(\ref{eq:EvTrench}) in the case
    of a trench of width $W=0.8$ for $\beta m g s = 5$. Left:
    $\beta P_{\mathrm{eff}} = 1$, Middle
    $\beta P_{\mathrm{eff}} = 500$.  Right: Diffusion coefficient
    (red) compared to the case of hard spheres in $2D$ for different values of $\beta m g s$.   Note that in
    the experiment of Ref.~\cite{Wei:2000}, the parameters give $\beta m g s \sim 200$, so it can be consider $2d$ in a wide range of density.}
  \label{fig:EVTrench}
\end{figure}

The procedure described in Section~\ref{sec:NumEval} can be directly adapted to obtain $\mu$ as a function of $P_{\mathrm{eff}}$. We then deduce $P_{\mathrm{eff}}(\rho)$ and $D(\rho)$ using the same Legendre transform~\eqref{eq:LegTrForMuRhoSM}.
The result is shown in Fig.~\ref{fig:EVTrench} in the case of hard spheres. The fact that the eigenvector is concentrated near $z=0$ seem to indicate that the particles remain at the bottom of the trench. We see that until $\rho \sim 0.9$, the case of the trench is identical to the $2D$ situation, but for larger values the curves deviate. Note that for these densities, the nearest-neighbour approximation no longer holds, since particles get on top of each other and the order is not conserved. We see from Fig.~\ref{fig:EVTrench} that if the gravity field is strong enough to confine the particle in a single layer, a trench is identical to a $2D$ channel. In the domain where $\rho > 1$, the gravitational field is not sufficient to ensure a strong enough confinement so the single-file constraint is not satisfied and our analytical calculations are not valid anymore.

\section{Consequences for single-file observables}

Having determined the fluctuating hydrodynamics equations~\eqref{eq:FlucHydroSM} and the equilibrium pressure in Section~\ref{sec:EqPr} which determine the transport coefficients~\eqref{eq:TrCoefsSM}, we can apply these results to study different observables. The results presented in this Section have already been derived in the literature~\cite{Krapivsky:2012,Krapivsky:2014,Krapivsky:2015a,Poncet:2021}, but we present here a derivation for completeness.

We consider two different types of initial conditions for the macroscopic density~\eqref{eq:DefMacroDens}:
\begin{enumerate}
\item[(i)] an "annealed" initial condition in which the system is initially at equilibrium at a mean density $\rb$. In particular, this implies for the macroscopic density~\eqref{eq:DefMacroDens}~\cite{Derrida:2025a}
\begin{equation}
	\label{eq:InitCorrel}
	\moy{ \rho(x,0) \rho(x',0) } - \rb^2
	= \Lambda^{-1} \frac{\sigma(\rb)}{2 D_S(\rb)} \delta(x-x')
	\:.
\end{equation}
\item[(ii)] a "quenched" initial condition in which the initial density is fixed at $\rho(x,0) = \rb$.
\end{enumerate}
We use here the terminology "annealed" and "quenched" used in the literature by analogy with disordered systems. 

Since $\Lambda \gg 1$, we can treat the noise term in the fluctuating hydrodynamics equation~\eqref{eq:FlucHydroSM} as a perturbation. We thus denote
\begin{equation}
	\label{eq:DeltaRho}
	\rho(x,t) = \rb + \Lambda^{-1/2} \delta \rho(x,t)
	\:.
\end{equation}
Inserting this definition into the evolution equation~\eqref{eq:FlucHydroSM} gives at leading order in $\Lambda^{-1/2}$,
\begin{equation}
	\partial_t \delta \rho = D_S(\rb) \partial_x^2 \delta \rho + \sqrt{\sigma(\rb)} \: \partial_x \eta
	\:.
\end{equation}
This is a linear diffusion equation for $\delta \rho$, which can be solved explicitly to give $\delta\rho(x,t)$ in terms of the noise $\eta$ and the initial condition $\rho(x,0)$,
\begin{align}
    \rho(x,t) 
    &= 
    \sqrt{\sigma(\bar\rho)} \int_{0}^t \dd t' \int_{-\infty}^\infty \dd x'
    \:
    K(x,t | x',0) \partial_{x'} \eta(x',t')
    + \int_{-\infty}^\infty \dd x'
    \: K(x,t | x',0) \rho(x,0)
    \:,
    \nonumber
    \\
    &=
    \sqrt{\sigma(\bar\rho)}  \partial_x \int_{0}^t \dd t' \int_{-\infty}^\infty \dd x'
    \:
    \eta(x',t')  K(x,t | x',0) 
    + \int_{-\infty}^\infty \dd x'
    \: K(x,t | x',0) \rho(x,0)
    \:,
\end{align}
where we introduced the heat kernel
\begin{equation}
	K(x,t | x',t')
	= \frac{\e^{- \frac{(x-x')^2}{4 D_S(\rb) (t-t')}}}{\sqrt{4 \pi D_S(\rb)(t-t')}}
	\:.
\end{equation}
Using that the noise correlations are given by~\eqref{eq:NoiseCorrel1D}, the initial correlations in the annealed case by~\eqref{eq:InitCorrel} and the two being independent, we get for the annealed case
\begin{equation}
	\label{eq:2ptCorrelAnnealed}
	\moy{\delta \rho(x,t) \: \delta \rho(x',t')}^{(a)}
	= \Lambda^{-1} \frac{\sigma(\rb)}{2 D_S(\rb)}
	\big[
		\Theta(t-t') K(x,t | x',t')
		+ \Theta(t'-t) K(x',t' | x,t)
	\big]
	\:,
\end{equation}
and for the quenched case
\begin{equation}
	\label{eq:2ptCorrelQuenched}
	\moy{\delta \rho(x,t) \: \delta \rho(x',t')}^{(q)}
	= \moy{\delta \rho(x,t) \: \delta \rho(x',t')}^{(a)}
	- \Lambda^{-1} \frac{\sigma(\rb)}{2 D_S(\rb)}
	K(x,t+t'|x',0)
	\:.
\end{equation}
From these expressions, we now deduce the fluctuations of various observables.

\subsection{For the integrated current}

We first consider the case of the current through a section $S$ of the channel, located for instance at $x=0$, integrated during a time $T$,
\begin{equation}
	\label{eq:DefQt}
	Q_T \equiv \int_0^T \dd t \int_S \dd^{d-1} \V{y} \:
	\jm((0,\V{y}),t) \cdot \V{e}_x
	\:.
\end{equation}
From the continuity equation~\eqref{eq:DKeq}, we can equivalently write it as
\begin{equation}
	Q_T = \int_0^\infty \dd x \int_S \dd^{d-1} \V{y} \big[
		\rhom((x,\V{y}),T) - \rhom((x,\V{y}),0)
	\big]
	\:.
\end{equation}
It can then be expressed in terms of the macroscopic density~\eqref{eq:DefMacroDens} as
\begin{equation}
	\label{eq:QtFromRhoMacro}
	Q_T = \Lambda \int_0^\infty \left[ 
		\rho(x,T/\Lambda^2) - \rho(x,0)
	\right] \dd x
	\:.
\end{equation}
We can now choose $\Lambda \propto \sqrt{T}$ so that $T/\Lambda^2$ remains of order $1$. The large observation time $T$ directly sets the scale $\Lambda$ of the macroscopic description.

Since the two point correlation of the density have an explicit form~(\ref{eq:2ptCorrelAnnealed},\ref{eq:2ptCorrelQuenched}), we directly deduce
\begin{equation}
	\label{eq:Qt2a}
	\moy{Q_T^2}^{(a)} 
	\underset{T \to \infty}{\simeq}	
	\frac{\sigma(\rb)}{\sqrt{\pi D_S(\rb)}} \sqrt{T}
	\:,
\end{equation}
\begin{equation}
	\label{eq:Qt2q}
	\moy{Q_T^2}^{(q)} 
	\underset{T \to \infty}{\simeq}	
	\frac{\sigma(\rb)}{\sqrt{2\pi D_S(\rb)}} \sqrt{T}
	\:.
\end{equation}
We recover the known expression of the fluctuations of the current $Q_T$~\cite{Krapivsky:2012}. Inserting the expressions of the transport coefficients~\eqref{eq:TrCoefsSM} and the pressure computed in Section~\ref{sec:EqPr} gives access to the exact long time behaviour of the integrated current in a channel.

\subsection{For the tracer's displacement}

Thanks to the single-file geometry, the displacement $X_T$ of a tracer during a time $T$ can be determined from the macroscopic density $\rho(x,t)$~\cite{Krapivsky:2014,Krapivsky:2015a}. Indeed, since the number of particles to the right of the tracer is conserved, we can write
\begin{equation}
	\int_0^{X_T} \dd x \int \dd^{d-1} \V{y} \: \rhom( (x,\V{y}), T)
	= \int_0^{\infty} \dd x \int \dd^{d-1} \V{y}
	\big[ \rhom( (x,\V{y}), T) - \rhom( (x,\V{y}), 0) \big]
	\:.
\end{equation}
Seen as an equation for $X_T$, it does not give a unique solution, since all the positions between the two neighbouring particles of the tracer are solution. This gives an error of order $1/\bar\rho$. At the macroscopic scale, the error $1/(\bar\rho \Lambda) \to 0$. Hence, the macroscopic version
\begin{equation}
	\int_0^{X_T/\Lambda}\rho( x, T/\Lambda^2) \dd x 
	= \int_0^{\infty}
	\big[ \rho( x, T/\Lambda^2) - \rhom( x, 0) \big] \dd x
\end{equation}
admits a single solution for $X_T$ for a given density. Inserting the expansion of the density~\eqref{eq:DeltaRho} for $\Lambda \gg 1$, we obtain
\begin{equation}
	\label{eq:RelXtQt}
	\rb \: X_T = Q_T
	\:
\end{equation}
Thus, we directly deduce from the fluctuations of the integrated current~(\ref{eq:Qt2a},\ref{eq:Qt2q}),
\begin{equation}
	\label{eq:Xt2}
	\moy{X_T^2}^{(a)} 
	\underset{T \to \infty}{\simeq}	
	\frac{\sigma(\rb)}{\rb^2 \sqrt{\pi D_S(\rb)}} \sqrt{T}
	\:,
	\qquad
	\moy{X_T^2}^{(q)} 
	\underset{T \to \infty}{\simeq}	
	\frac{\sigma(\rb)}{\rb^2 \sqrt{2\pi D_S(\rb)}} \sqrt{T}
	\:.
\end{equation}
We recover the known results~\cite{Krapivsky:2014,Krapivsky:2015a}. Note that the relation~\eqref{eq:RelXtQt} is only valid for the fluctuations. The distributions of these quantities are different, and higher order cumulants of these observables differ (see for instance the fourth cumulants derived in Ref.~\cite{Grabsch:2024b}).

\subsection{Higher order cumulants}

\begin{figure}
    \centering
    \includegraphics[width=0.4\textwidth]{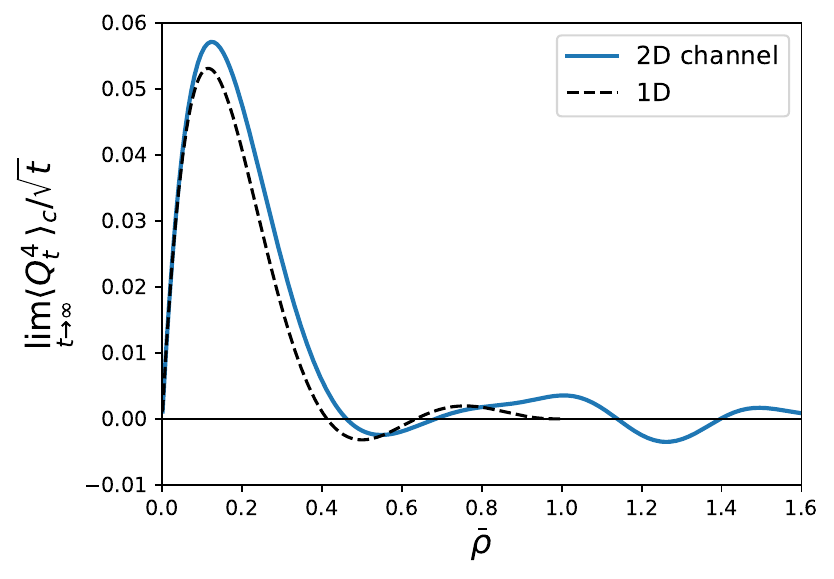}
    \includegraphics[width=0.4\textwidth]{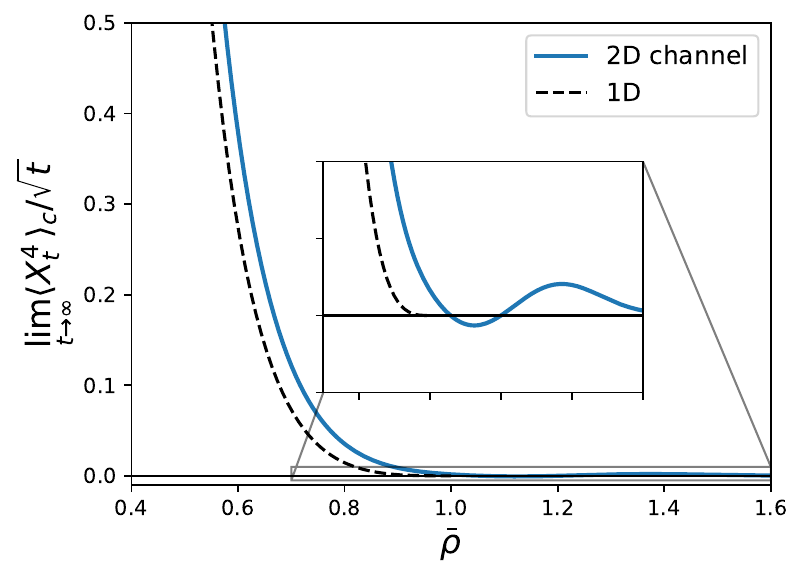}
    \caption{Fourth cumulants $\moy{Q_t^4}_c$ and $\moy{X_t^4}_c$ as a function of the mean density $\bar\rho$, for an annealed initial condition. We have considered a $2D$ channel of width $1.86$ filled with particles of size $s=1$.}
    \label{fig:Kappa4}
\end{figure}

Higher order cumulants can be computed from the fluctuating hydrodynamics equation~\eqref{eq:FlucHydroSM}, although the procedure becomes more cumbersome. For instance, the fourth cumulants of $X_t$ and $Q_t$, which measure the deviation from the Gaussian behavior, have been computed explicitly for a general system~\cite{Grabsch:2024b}. The expressions are rather lengthy, so we do not reproduce them here. Plots of the density dependence of these cumulants are given in Fig.~\ref{fig:Kappa4}.

\subsection{Coupling with the density}

Proceeding similarly, we compute from the two-point correlations in the annealed case~\eqref{eq:2ptCorrelAnnealed},
\begin{equation}
    \moy{Q_t \: \rho(x,t)}^{(a)}
    = \frac{\sigma(\rb)}{4 D_S(\rb)} \sg{x} \erfc \left(
    \frac{|x|}{\sqrt{4 D_S(\rb) t}}
    \right)
    \:,
\end{equation}
\begin{equation}
    \moy{X_t \: \rho(X_t + x,t)}^{(a)}
    = \frac{\sigma(\rb)}{4 \rb D_S(\rb)} \sg{x} \erfc \left(
    \frac{|x|}{\sqrt{4 D_S(\rb) t}}
    \right)
    \:,
\end{equation}
Recovering the known expressions~\cite{Poncet:2021,Grabsch:2022,Grabsch:2023}.
Similarly, in the quenched case, we get
\begin{equation}
    \moy{Q_t \: \rho(x,t)}^{(q)}
    = \frac{\sigma(\rb)}{4 D(\rb)} \sg{x} \erfc \left(
    \frac{|x|}{\sqrt{8 D(\rb) t}}
    \right)
    \:,
\end{equation}
\begin{equation}
    \moy{X_t \: \rho(X_t + x,t)}^{(q)}
    = \frac{\sigma(\rb)}{4 \rb D(\rb)} \sg{x} \erfc \left(
    \frac{|x|}{\sqrt{8 D(\rb) t}}
    \right)
    \:.
\end{equation}

\section{Extension to underdamped dynamics}

The formalism presented in the article directly extends beyond the case of Brownian (overdamped Langevin) particle to describe the large scale behaviour of underdamped Langevin particles. This is the goal of this Section.

We now consider that the particles evolve according to the coupled Langevin equations
\begin{equation}
	m \frac{\dd^2 \V{r}_n}{\dd t^2}
	= - \gamma  \frac{\dd \V{r}_n}{\dd t}
	- \sum_{m \neq n} \nabla V(\V{r}_n-\V{r}_{m})
	+\sqrt{2\gamma k_{\rm B}T}\,\V{\eta}_n(t)
	\:,
	\label{eq:langevin_channel_damp}
\end{equation}
instead of~\eqref{eq:langevin_channel_all} with the same Gaussian white noises $\V{\eta}_n$ as before. The parameter $\gamma$ is the damping factor.

In this case, the microscopic density of particles obey the stochastic equation~\cite{Nakamura:2009}
\begin{subequations}
  \label{eq:DKunderdamped}
  \begin{align}
    \label{eq:ConsMicro}
    \partial_t \rho_0
    &= - \V{\nabla} \cdot \V{j}_0
    \\
    \label{eq:jMicro}
    \partial_t j_0^i
    &=
      - \gamma j_0^i
      - \sum_j \sum_n \frac{p_n^i p_n^j}{m} \nabla_i \delta^{(d)}(\V{r} - \V{r}_n(t))
      - \rho_0 \nabla_i (V_0 \star \rho_0)
      + \sqrt{2 k_{\mathrm{B}}T \gamma  \rho_0} \: \eta_i
      \:,
  \end{align}
\end{subequations}
where $\V{p}_n$ is the momentum of particle $n$, the convolution is still defined by~\eqref{eq:DefConvol}. We again define the macroscopic fields by~(\ref{eq:DefMacroDens},\ref{eq:DefMacroCurr}). Most of the terms in these equations can be coarse-grained as in Section~\ref{sec:FlucHydro} above. The only new term is the one involving the momenta in~\eqref{eq:jMicro}. The coarse-graining of this term is done in~\cite{Das:2013} in the unconfined geometry. Assuming this same derivation holds for the channel geometry gives
\begin{equation}
  \frac{\Lambda}{\ell} \int_0^\ell \dd x' \int \dd^{d-1} \V{y} \:
  \sum_j \sum_{n=1}^N \frac{p_n^x p_n^j}{m} \nabla_j [\delta(\Lambda x + x' - x_n(\Lambda^2 t)) \delta^{(d-1)}(\V{y} - \V{y}_n(\Lambda^2 t))]
  =  k_{\mathrm{B}}T \partial_x \rho
  + \Lambda^{-2} \nabla_x \left[ \frac{j^2}{\rho} \right]
  \:.
\end{equation}
Therefore, combining this expression with the results of Section~\ref{sec:FlucHydro} for the other terms, Eq.~\eqref{eq:jMicro} becomes
\begin{equation}
	\Lambda^{-2} \partial_t j
	= - \gamma j
	- \partial_x P_{\mathrm{eff}}(\rho)
	- \Lambda^{-2}  \nabla_x \left[ \frac{j^2}{\rho} \right]
	- \sqrt{\frac{2 k_{\mathrm{B}} T \gamma \rho}{\Lambda}} \eta
	\:.
\end{equation}
Keeping only the two leading terms for $\Lambda \gg 1$, we obtain the same expression for the current as in the overdamped case
\begin{equation}
	j = - \mu_0 P_{\mathrm{eff}}'(\rho) \partial_x \rho
	- \sqrt{\frac{2 \mu_0 k_{\mathrm{B}} T\rho}{\Lambda}} \eta
	\:,
\end{equation}
with $\mu_0 = \gamma^{-1}$. This is expected since at macroscopic (i.e. large) times, the damping has occurred and the dynamics is the same as for Brownian particle.

\section{Numerical simulations}

The simulations are performed using the LAMMPS software~\cite{Thompson:2022}. Most simulations are performed in a periodic channel of width $0.86$ and of different length $L$ for $N=800$ particles. The density of particles is $\rb = N/L$. The hardcore interaction is approximated by a WCA potential, using the LAMMPS pair potential \texttt{lj/cut}. We start with a fixed initial conditions in which the particles are equally spaced and their initial velocities is set to zero (this realizes the ``quenched'' initial condition and thus avoids issues related to thermalization of the initial state). We simulate the Langevin equations~\eqref{eq:langevin_channel_damp} with $\gamma = m/\texttt{damp}$, with the LAMMPS parameter $\texttt{damp}$ set to $m$ so that $\gamma = 1/\mu_0 = 1$. In most cases we use $m = 0.1$ to simulate overdamped particles. The simulations are performed up to time $T=500$  with a timestep $\dd t = 0.001$. The observables are computed every $\delta t = 25$. Within each simulation, the position $X_t$ of the tracer, as well as $X_t^2$, are averaged over all the particles (each one playing the role of the tracer). The integrated current $Q_t$ (and $Q_t^2$) is also averaged within every simulation at $50$ different measuring points equally spaced along the system. These results are then averaged over $200$ realizations and 95\% confidence intervals are obtained from a bootstrap resampling of these realizations.

\begin{figure}
    \centering
    \includegraphics[width=0.45\textwidth]{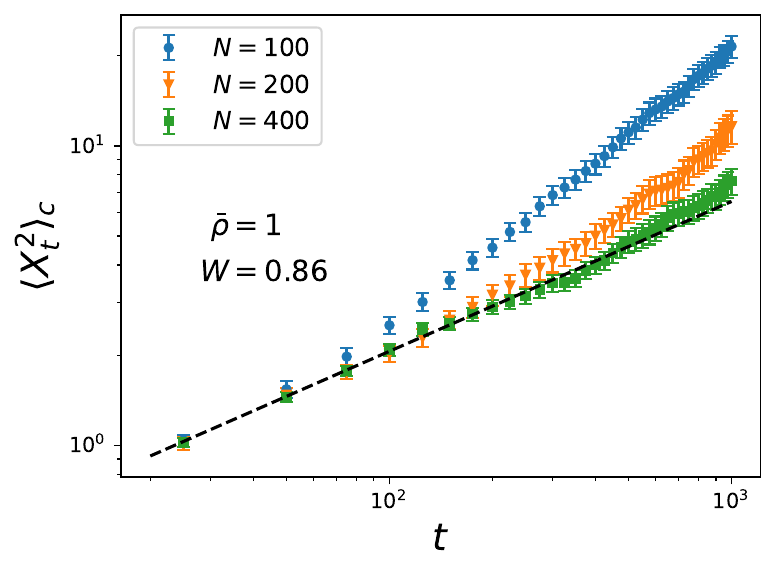}
    \caption{Finite size effects on the mean squared displacement of a tracer. The points are obtained from numerical simulations with LAMMPS of particles interacting via a WCA potential in a $2D$ channel of width $1.86$ at mean density $\bar\rho = 1$ for different numbers of particles, with $m = 0.1$, $\gamma = 1/\mu_0 = 1$, $T = 1$. The results are averaged over $500$ simulations and the error bars are obtained from a bootstrap resampling of these independent realizations. The dashed line is the theoretical single-file prediction in an infinite system.}
    \label{fig:FiniteSizeEffects}
\end{figure}

The amplitudes of the fluctuations of $X_t$ and $Q_t$ are obtained by fitting $\ln \moy{X_t^2}$ and $\ln \moy{Q_t^2}$ with $a + \frac{1}{2}\ln t$, over the interval $[100,500]$. These values have been chosen to ensure that:
\begin{enumerate}[label=(\roman*)]
    \item The crossover from the normal diffusion at short time to the single-file behavior is reached (i.e. for both $t \gg 1/(D(\bar\rho) \bar\rho^2)$ to ensure interaction between neighboring particles and $t \gg m/\gamma$ to reach the diffusive behavior). With the parameters used, both are satisfied for $t > 10$. We start from $t=100$ to ensure that the single-file regime has been reached, but a smaller value could be used. Note that increasing the density reduces the crossover time and makes the single-file regime emerge at shorter times.
    \item Finite size effects are subdominant. These become important once the diffusion of the density has reached the periodic boundary. This occurs at a time $t \sim L^2/D(\bar\rho) = N^2/(D(\bar\rho) \bar\rho^2)$. After this time, the observables display a linear behavior in time. See Fig.~\ref{fig:FiniteSizeEffects} for an illustration of these finite size effects.
\end{enumerate}
The error bars are again obtained from a bootstrap resampling of the realizations.

%